\documentclass[twocolumn,amsmath,amssymb,aps,prx]{revtex4-1}

\usepackage{blindtext}
\usepackage{bm}
\usepackage{graphicx,epsfig,epstopdf}
\usepackage{amsmath}
\usepackage{color}
\usepackage{subfigure}
\usepackage{array}
\usepackage{tabularx}
\usepackage{multirow}
\usepackage[export]{adjustbox}
\newcolumntype{L}[1]{>{\raggedright\arraybackslash}p{#1}}
\newcolumntype{C}[1]{>{\centering\arraybackslash}p{#1}}
\newcolumntype{M}[1]{>{\centering\arraybackslash}m{#1}}

\def\_#1{{\bf #1}}
\def\.{\cdot}

\def\Re{{\rm Re\mit}}

\begin{document}

\title{Multiphysics  metamirrors for simultaneous manipulations of acoustic and electromagnetic waves}

\author{
 Ana~D\'{i}az-Rubio  and Sergei~Tretyakov
}
 
\affiliation{Department of Electronics and Nanoengineering, Aalto University, P.~O.~Box~15500, FI-00076 Aalto, Finland}

\begin{abstract} 
Metasurfaces have shown unprecedented possibilities for wavefront manipulation of waves. The research efforts have been focused on the development of metasurfaces that perform a specific functionality for waves of one physical nature, for example, for electromagnetic waves. In this work, we propose the use of power-flow conformal metamirrors for creation of  multiphysics devices which can simultaneously control waves of different nature. In particular, we introduce metasurface devices which perform specified operations on both electromagnetic and acoustic waves at the same time. Using a purely analytical model based on surface impedances, we introduce metasurfaces that perform the same functionality for electromagnetic and acoustic waves and, even more challenging, different functionalities for electromagnetics and acoustics.  We provide realistic topologies for practical implementations of proposed metasurfaces and confirm the results with numerical simulations. 
\end{abstract}

\maketitle
\section{Introduction}

Over the last few years, great efforts have been concentrated on studying new possibilities for controlling wave propagation using microstructured surfaces, so-called metasurfaces. 
Research on metasurfaces covers aspects from the theory of wave propagation, fabrication techniques, and even different disciplines or application fields such as acoustics \cite{Review_AC} or electromagnetism \cite{Review_EM} \cite{Quevedo_Roadmap_2019}. Metasurfaces have shown great potential for manipulating waves both in transmission and reflection. In particular, metasurfaces controlling reflection, also known as metamirrors, have been developed for realizations  of retroreflectors, anomalous reflectors, beam splitters, and focusing devices \cite{Estakhri_Wavefront_2016, Diaz_From_2017, Epstein_Unveiling_2017, Radi_Metagrating_2018, DoHoon_Lossless_2018, Asadchy_Flat_2017, Diaz_Acoustic_2017, Torrent_2018,Diaz_Power_2019, Shen_Surface_2018}.

In the design of reflective metasurfaces, one can distinguish two different types of design strategies. On the one hand, there are examples of several \textit{non-local design approaches} where the properties on each point of the metasurface depend not only on the local field at this specific point but also on the fields in the vicinity of  this point \cite{ Diaz_From_2017,  Radi_Metagrating_2018, Diaz_Acoustic_2017, Popov_Controlling_2018, Torrent_2018,Epstein_Unveiling_2017}. Exploiting non-locality it is possible to reduce the number of meta-atoms required for the implementation. However, the corresponding design approaches are complicated especially for complex functionalities such as focusing. On the other hand, using \textit{local design approaches} the properties of the metasurface at each point can be uniquely characterized by response on the local fields at this point \cite{  DoHoon_Lossless_2018, Diaz_Power_2019, Shen_Surface_2018}. In this case, the metasurface can be characterized by local parameters such as the surface impedance.  The main advantage of these methods is that knowing the desired performance of the metasurface, the design of metasurfaces becomes systematic and in most of the cases it does not require  numerical optimizations. In  recent work \cite{Diaz_Power_2019}  it was demonstrated that 
local design of metasurfaces is possible by proper engineering the shape and the surface impedance according to the power-flow distribution of the required set of waves.

In this work, we will show that exploiting the properties of local metasurfaces one can design multiphysics devices which can simultaneously control waves of different nature. This becomes possible because power flow-conformal metasurfaces can be realized as arrays of small (subwavelength) meta-atoms acting as phase-shifting elements \cite{Diaz_Power_2019}. Here we exploit a possibility to create meta-atoms which can provide controllable phase shifts for both electromagnetic waves and for sound, which open up a way towards creation of high-performance multi-functional multi-physics metasurface devices.


\begin{figure}[b]
	\subfigure[]{\includegraphics[width=1\columnwidth,left]{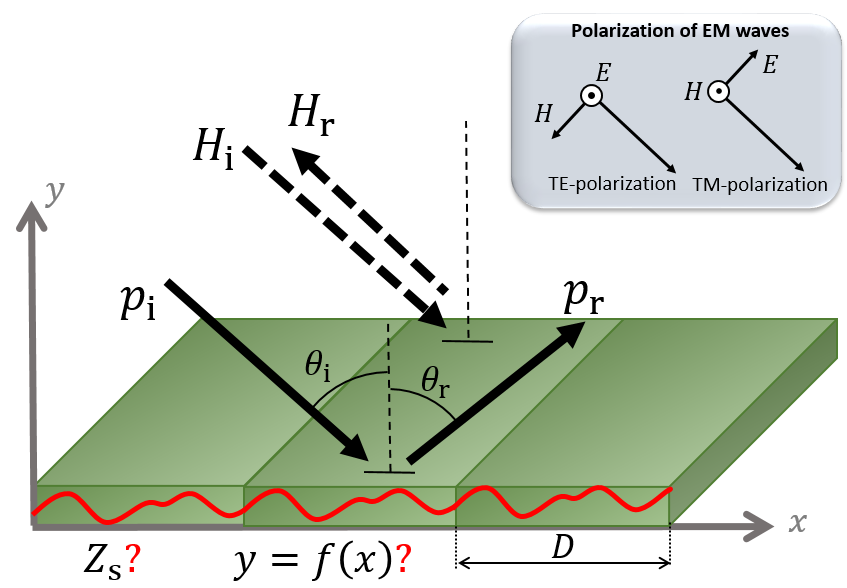}\label{fig:Scheme1}}
	\caption{ Schematic representation of a multiphysics metamirror.  }
	\label{fig:Schemes}
\end{figure}

Although acoustic and electromagnetic waves  obey different wave propagation equations that, respectively, govern sound pressure and electromagnetic field distributions, there exists a direct analogy that allows to create multi-physics structures \cite{Carbonell_Multidisciplinary_2011}.
In particular, we study the analogy of impedance-based models for electromagnetic and acoustic reflectors. 
In this paper, we exploit this analogy and present a direct and comparative approach for the design and implementation of power flow-conformal metamirrors for both acoustic and electromagnetic fields.
We demonstrate that using power flow-conformal metamirrors and the local-parameter approach, the design of such multiphysics devices becomes systematic and straightforward.  
Several example functionalities are analyzed and characterized in terms of the shape and surface impedance of the metamirror, as well as topologies of physical implementations. 
The study shows similarities and differences between both wave problems (acoustic and electromagnetic) that have to be considered in the final implementation of multiphysics platforms.

The paper is organized as follows.
First, in Section~II, we present the theory for the design of power flow-conformal  metamirrors paying special attention to the similarities and differences between the two domains of physics.
In particular, the study will be focused on manipulation of two plane waves (retroreflectors and anomalous reflectors).
This allows us to streamline the design of metamirrors that allow simultaneous control of both acoustic and electromagnetic responses. 
Next, in Section~III, we discuss practical aspects related to actual implementations of  multi-physics metamirrors. 
Finally, some conclusions summarize the main findings
of this work.

\section{General theoretical framework}



In this section, we present the theory of power flow-conformal metamirrors and highlight the differences between acoustic and electromagnetic  devices for shaping reflected waves.
More specifically, we will discuss the design of reflective surfaces for reflecting an incident plane wave into a desired direction. The key parameter to be considered in the solutions of both problems is the surface impedance, which can be defined for both electromagnetic and acoustic waves.

Let us consider  a curved impenetrable boundary lying on  the $xz$-plane whose shape is defined by the function $y_{\rm c}=f(x)$ (for simplicity we assume that the surface profile is uniform along the orthogonal direction $z$). For acoustic waves, the surface impedance $Z_{\rm s}(x)$ is defined as a relation between the pressure $p$   and velocity ${\bf v}$ fields: $Z_{\rm s}(x)\hat{\bf n}\cdot{\bf v}(x,y_{\rm c})=-p(x,y_{\rm c})$. Here, $\hat{\bf n}$ is the unit  vector normal to the surface of the metamirror at point $x$. 
For electromagnetic waves, the surface impedance is defined as a relation between the the tangential to the surface components of the electric and magnetic fields, ${\bf E}_t$ and ${\bf H}_t$, as $Z_{\rm s}(x)\hat{\bf n}\times {\bf H}_t(x,y_{\rm c})={\bf E}_t(x,y_{\rm c})$. 

The wave fields above a perfect anomalously reflecting metamirror  are sums of fields of only two plane waves: one is the incident plane wave and the other is the reflected plane wave, which propagates in the desired, freely chosen direction. 
In order to find the optimal surface profile, $y_{\rm c}=g(x)$, which ensures the desired functionalities of the metamirror, the first step is to analyze the distribution of power flow in this set of two plane waves \cite{Diaz_Power_2019}. In what follows, we will analyse the power flow  for both electromagnetic and acoustic problems.

\textit{\underline{Acoustic metamirror}:} The wave vectors of the waves can be defined as $\bm{k}_{\rm i,r}=k_{{\rm i,r}}^{(x)}\hat{\bf x}+k_{{\rm i,r}}^{(y)}\hat{\bf y})$, where the subscript $\rm i$($\rm r$) denotes the incident (reflected) plane wave. In the most general case, we can define the incident wavevector as $\bm{k}_{\rm i}=k(\sin{\theta_{\rm i}}\hat{\bf x}-\cos{\theta_{\rm i}}\hat{\bf y})$, where $\theta_{\rm i}$ is the incident angle (anti-clockwise definition)  and  $k=\omega/c_{\rm ac}$ is the wavenumber of sound waves in the background medium at the frequency of interest ($c_{\rm ac}$ is the speed of sound). Following the same notation, the wavevector of the reflected wave is defined as  $\bm{k}_{\rm r}=k(\sin{\theta_{\rm r}}\hat{\bf x}+\cos{\theta_{\rm r}}\hat{\bf y})$ with $\theta_{\rm r}$ being the reflection angle (clockwise definition). A schematic representation of this scenario is shown  in Fig.~\ref{fig:Schemes}.
Under the $e^{j\omega t}$ time convention, the field can expressed as follows:
\begin{eqnarray}
p(x,y)=p_{\rm i}e^{-j\bm{k}_{\rm i}\cdot \bm{r}}+p_{\rm r}e^{-j\bm{k}_{\rm r}\cdot \bm{r}}\\
v_x(x,y)=\frac{1}{\omega \rho}\left[p_{\rm i}k_{{\rm i}}^{(x)}e^{-j\bm{k}_{\rm i}\cdot \bm{r}}+p_{\rm r}k_{{\rm r}}^{(x)}e^{-j\bm{k}_{\rm r}\cdot \bm{r}}\right]\\
v_y(x,y)=\frac{1}{\omega \rho}\left[p_{\rm i}k_{{\rm i}}^{(y)}e^{-j\bm{k}_{\rm i}\cdot \bm{r}}+p_{\rm r}k_{{\rm r}}^{(y)}e^{-j\bm{k}_{\rm r}\cdot \bm{r}}\right]
\end{eqnarray}
where $p_{\rm i}=\vert p_{\rm i}\vert e^{j\phi_{\rm i}}$ and $p_{\rm r}=\vert p_{\rm r}\vert e^{j\phi_{\rm r}}$ are the complex amplitudes of the incident and reflected plane waves, $\bm{r}=x\hat{\bf x}+y\hat{\bf y}$ is the position vector, $\omega$ is the angular frequency, and $\rho$ is the density of the background media.  From the definition of the pressure and velocity vector, it is straightforward to derive the intensity vector of the superposition of both incident and reflected plane waves as $\bm{I}(x,y)=\frac{1}{2}\Re{(p \bm{v}^*)}=I_x(x,y)\hat{\bf x}+I_y(x,y)\hat{\bf y}$. The analytical expression of  the flow of power  carried by two arbitrary plane waves reads  
\begin{equation}
\begin{split}
I_x(x,y)&= I_0\left[\vert p_{\rm i}\vert^2 k_{{\rm i}}^{(x)} +\vert p_{\rm r}\vert^2 k_{{\rm r}}^{(x)}\right]\\
& +I_0\vert p_{\rm i}\vert\vert p_{\rm r}\vert(k_{{\rm i}}^{(x)}+k_{{\rm r}}^{(x)})\cos(\Delta\bm{k}\cdot\bm{r}+\Delta\phi)
\end{split}\label{eq:Ix_2PW}
\end{equation}
and
\begin{equation}
\begin{split}
I_y(x,y)&= I_0\left[\vert p_{\rm i}\vert^2 k_{{\rm i}}^{(y)} +\vert p_{\rm r}\vert^2 k_{{\rm r}}^{(y)}\right]\\
& +I_0\left[\vert p_{\rm i}\vert\vert p_{\rm r}\vert(k_{{\rm i}}^{(y)}+k_{{\rm r}}^{(y)})\cos(\Delta\bm{k}\cdot\bm{r}+\Delta\phi)\right]
\end{split}, \label{eq:Iy_2PW}
\end{equation}
where $I_0=1/2\omega\rho$, $\Delta\bm{k}=\bm{k}_{\rm r}-\bm{k}_{\rm i}$, and $\Delta\phi=\phi_{\rm i}-\phi_{\rm r}$.

\begin{figure*}
\minipage{0.32\textwidth}
	\subfigure[]{\includegraphics[width=1\linewidth,left]{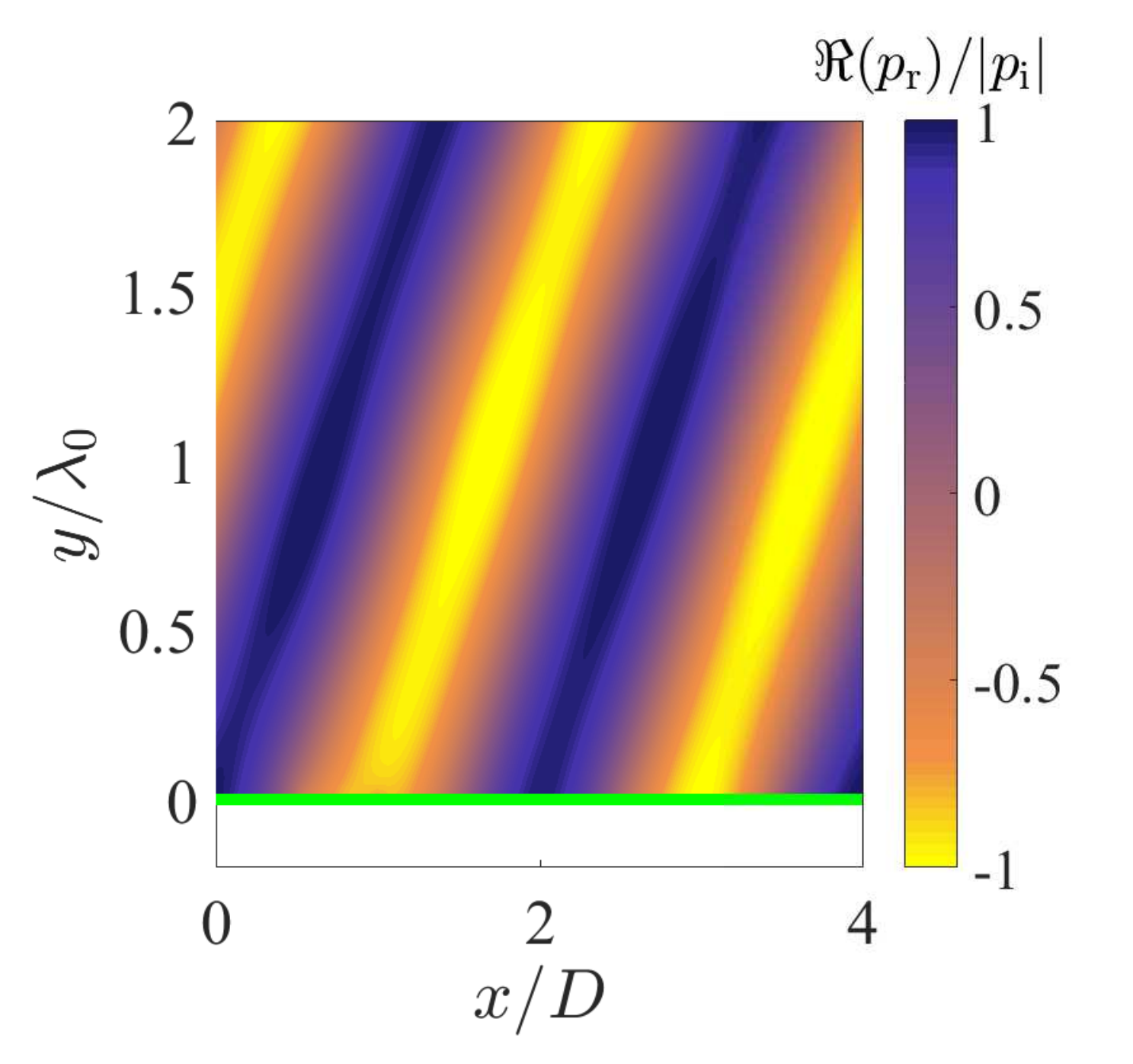}\label{fig:RetroreflectorA}}
	\subfigure[]{\includegraphics[width=0.8\linewidth,left]{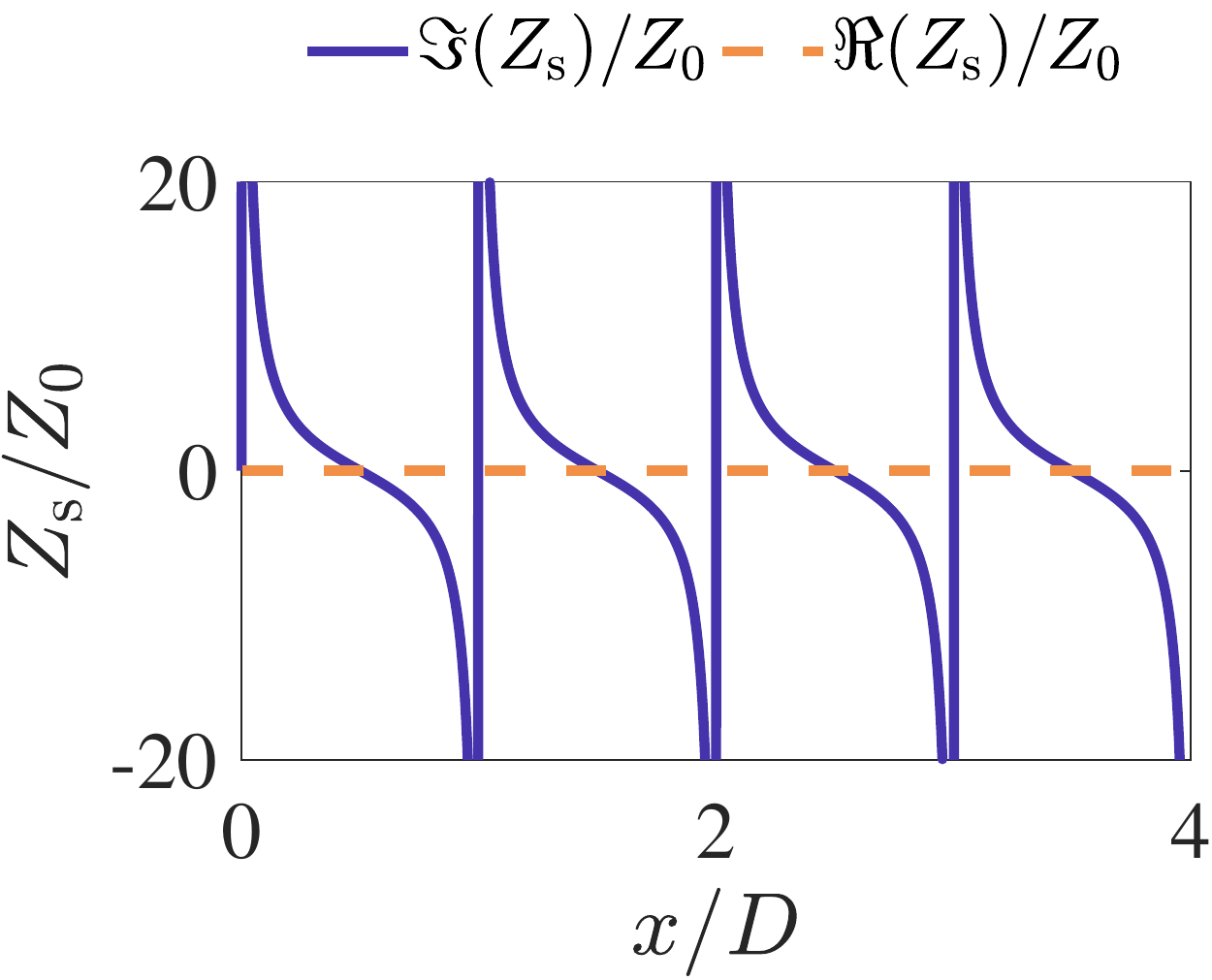}\label{fig:RetroreflectorB}}
\endminipage\hfill
\minipage{0.32\textwidth}
\subfigure[]{\includegraphics[width=1\linewidth,left]{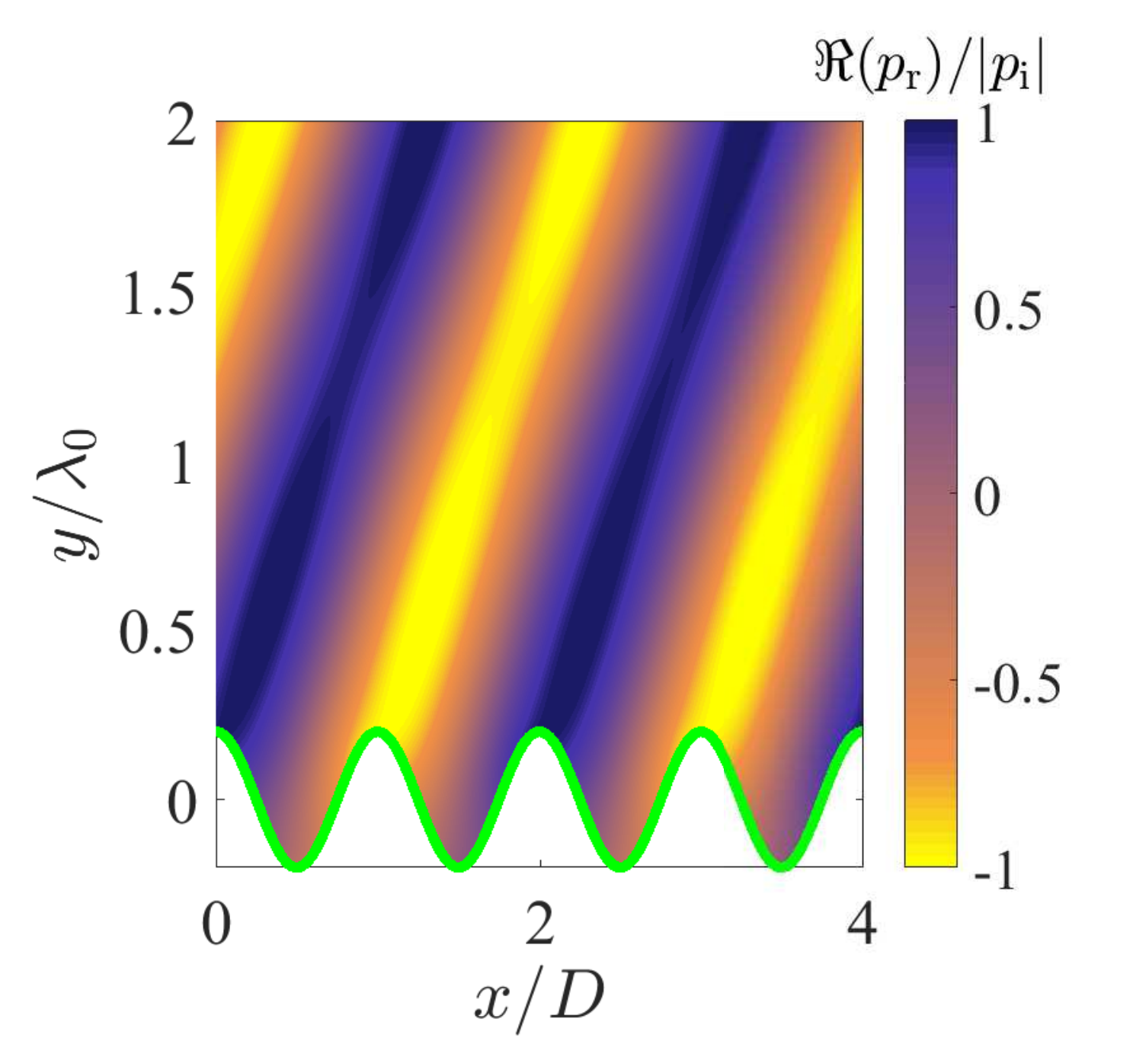}\label{fig:RetroreflectorC}}
\subfigure[]{\includegraphics[width=0.8\linewidth,left]{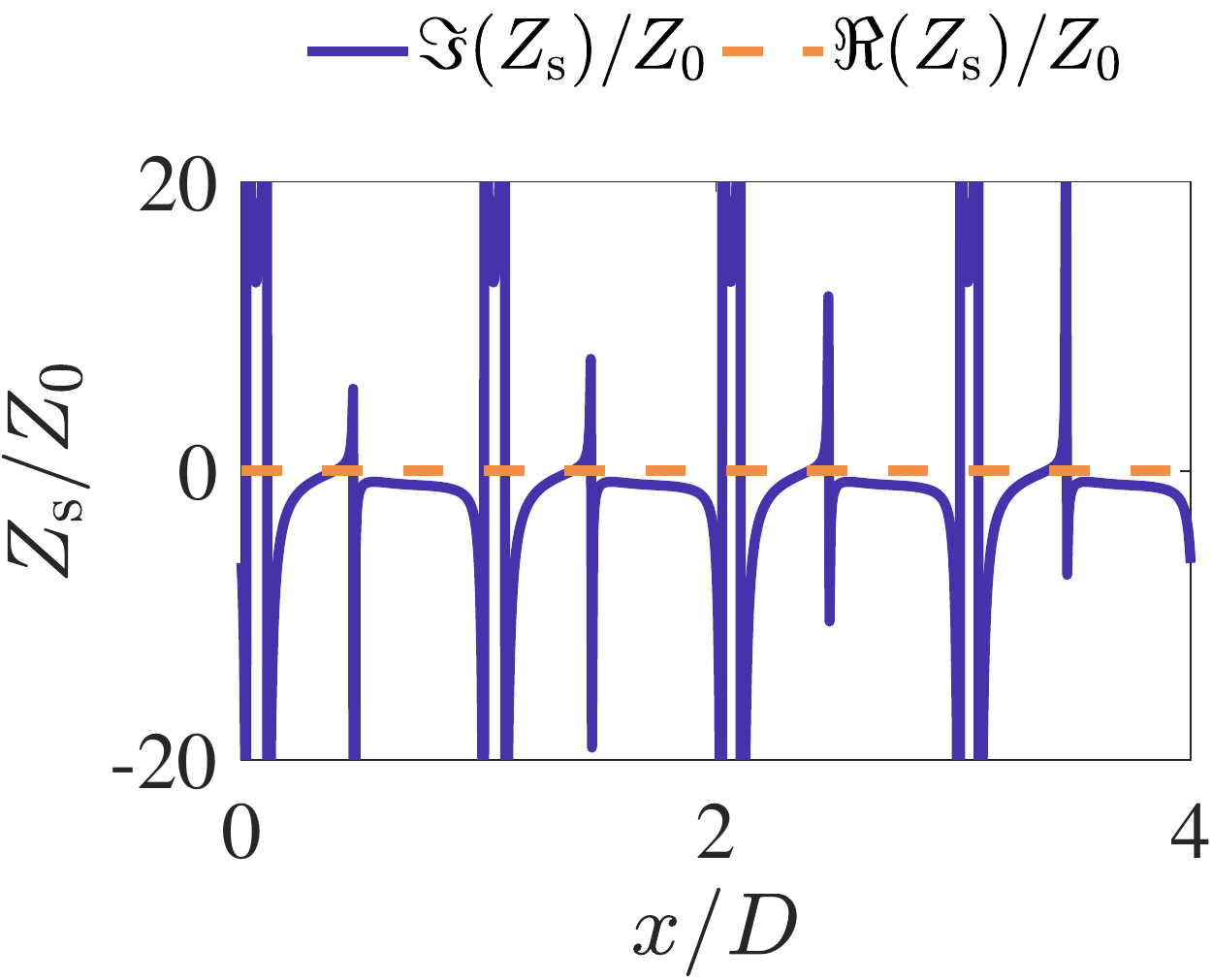}\label{fig:RetroreflectorD}}
\endminipage\hfill
\minipage{0.32\textwidth}
\subfigure[]{\includegraphics[width=1\linewidth,left]{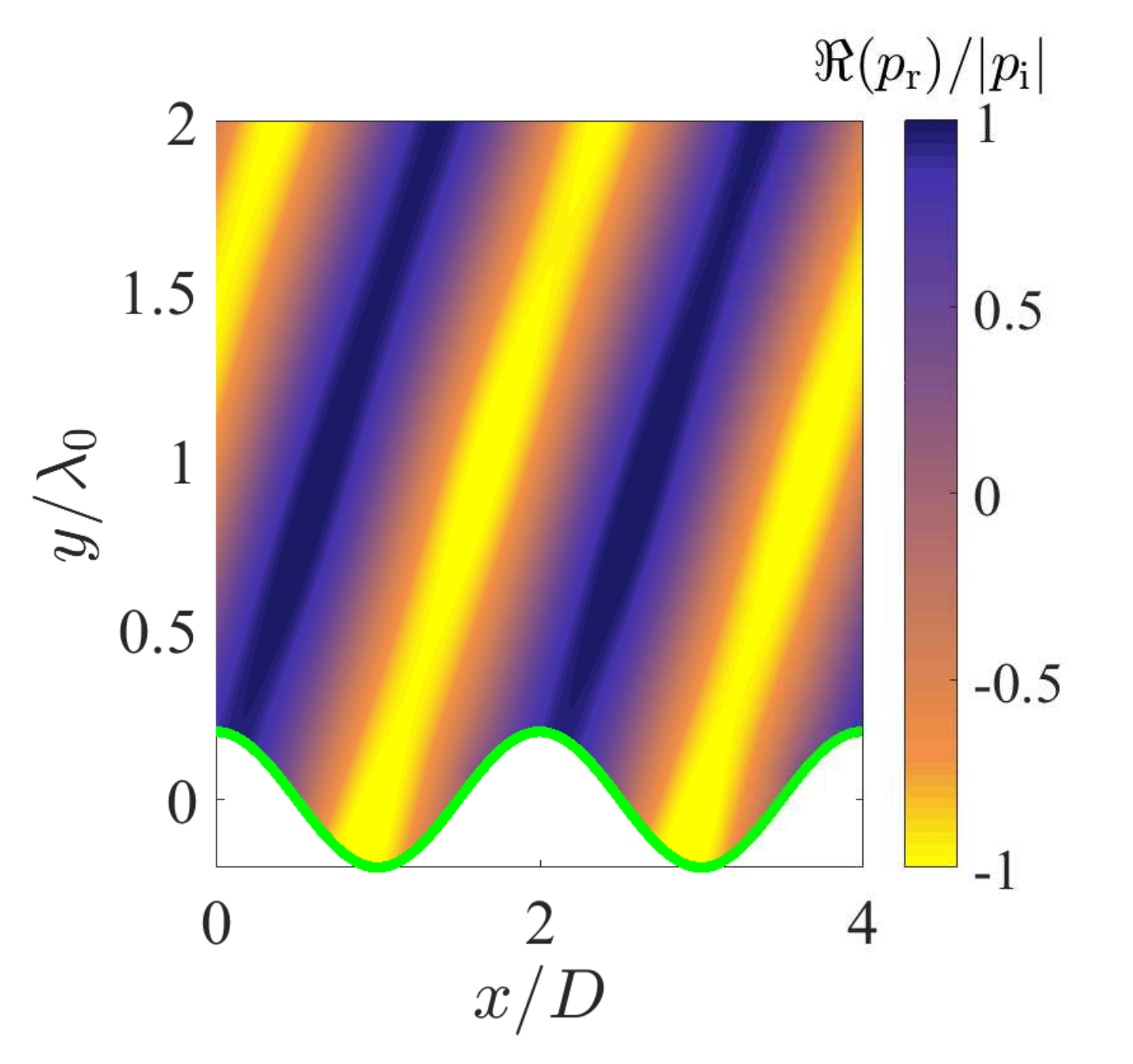}\label{fig:RetroreflectorE}}
\subfigure[]{\includegraphics[width=0.8\linewidth,left]{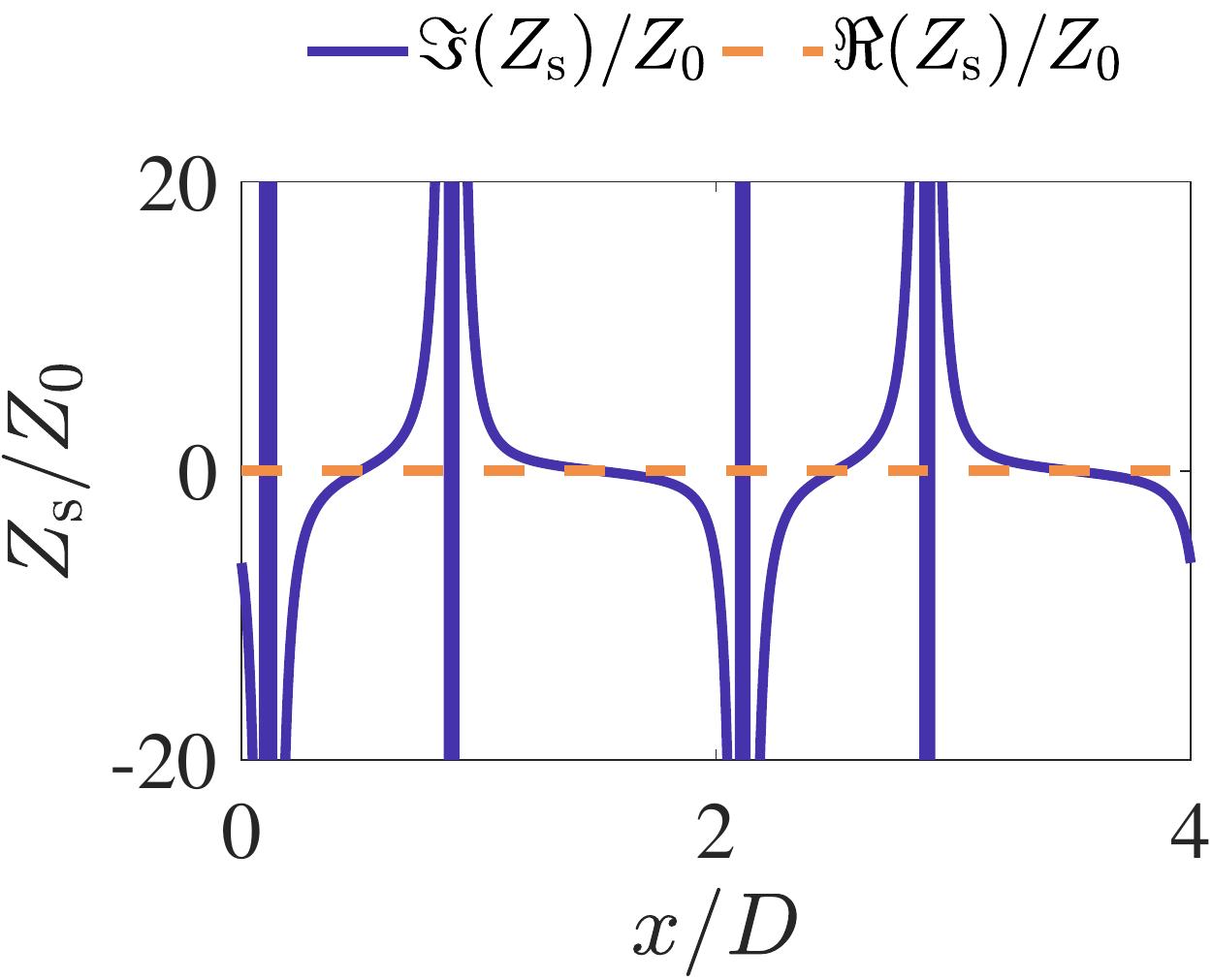}\label{fig:RetroreflectorF}}
\endminipage\hfill
	\caption{Acoustic power flow-conformal retroreflectors for $\theta_{\rm i}=70^\circ$.  Reflected pressure field (a) and normalized surface impedance (b) for the flat retroreflector.  Reflected pressure field (c) and normalized surface impedance (d) for the retroreflector with  cosine profile  $y_{\rm c}=\frac{\lambda}{5}\cos{(2\pi x/D)}$.  Reflected pressure field (e) and normalized surface impedance (f) for the retroreflector with  cosine profile $y_{\rm c}=\frac{\lambda}{5}\cos{(\pi x/D)}$. }
	\label{fig:Retroreflector}
\end{figure*}

\textit{\underline{Electromagnetic metamirror}:} The analysis of the power-flow distribution for electromagnetic waves can be done following a similar approach.
Because of the vectorial nature of the problem, here we should distinguish between TM-polarized ($H_z$, $E_x$, and $E_y$) and TE-polarized ($E_z$, $H_x$, and $H_y$) waves. Notice that in this work we do not consider any change in the polarization state and assume that both incident and reflected waves have the same polarization.  As it was shown in \cite{Diaz_Power_2019}, the power-flow distribution in the electromagnetic counterpart defined by the Poynting vector $\bm{S}=\frac{1}{2}\Re{(\bm{E}\times\bm{H}^*)}=S_x(x,y)\hat{\bf x}+S_y(x,y)\hat{\bf y}$ is given by similar 
expressions as in Eqs.~(\ref{eq:Ix_2PW}) and (\ref{eq:Iy_2PW}). There is a clear analogy between the two physical problems:  For TM-polarized waves, the amplitude of the pressure fields is analogous to the amplitude of magnetic fields ($p_{\rm i,r}\xrightarrow{}H_{\rm i,r}$) and the density of the background media to the permittivity ($\rho\xrightarrow{}\varepsilon_0$). For TE-polarized waves the amplitude of electric fields plays the role of the amplitude of the pressure fields  ($p_{\rm i,r}\xrightarrow{}E_{\rm i,r}$) and the density of the background media is analogous to the permeability ($\rho\xrightarrow{}\mu_0$).

From the analysis of the power-flow density distribution  we can see that the power-flow distribution mainly depends on the amplitude of the plane waves ($\vert p_{\rm i}\vert$ and ($\vert p_{\rm r}\vert$) and the direction of propagation ($\bm{k}_{\rm i}$ and $\bm{k}_{\rm r}$). We can distinguish three different scenarios depending on the relations between these parameters: (i)  When $\vert p_{\rm i}\vert=\vert p_{\rm r}\vert$ and $\bm{k}_{\rm r}=-\bm{k}_{\rm i}$, the wave is reflected back towards the source, and we call it the retroreflection scenario. (ii)  Reflections with $\vert p_{\rm i}\vert=\vert p_{\rm r}\vert$ but $\bm{k}_{\rm r}\ne \bm{k}_{\rm i}$ can be realized if the wave reflects specularly. (iii)  Finally, full power reflection is possible with $\vert p_{\rm i}\vert\ne \vert p_{\rm r}\vert$ and $\bm{k}_{\rm r}\ne \bm{k}_{\rm i}$. This case is usually called anomalous reflection. In the following, we will consider in detail multiphysics realizations of all these scenarios.

\begin{figure*}
	
	\minipage{0.32\textwidth}
	\subfigure[]{\includegraphics[width=1\linewidth,left]{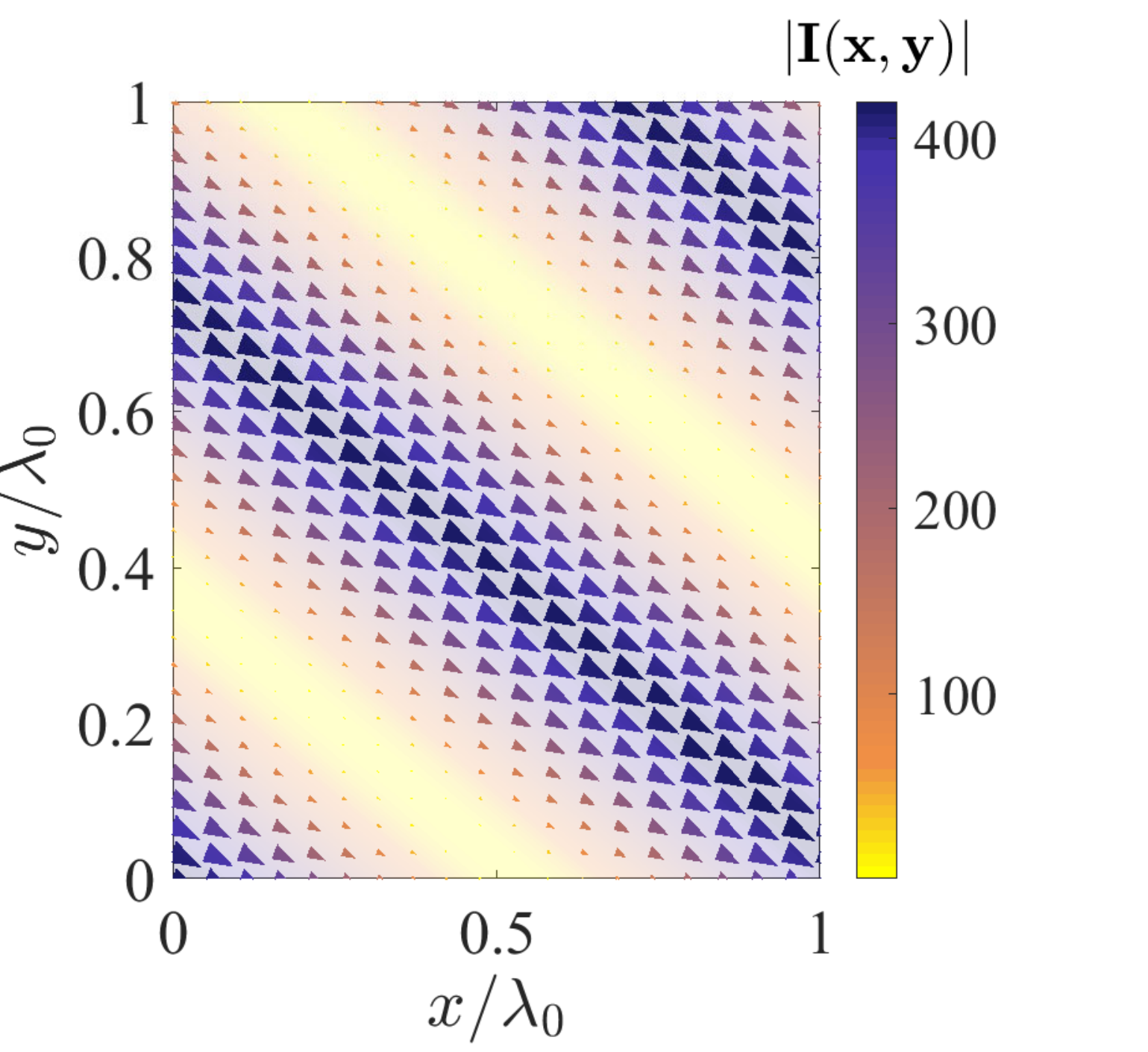}\label{fig:SpecularreflectionA}}
	\endminipage\hfill
	\minipage{0.32\textwidth}
	\subfigure[]{\includegraphics[width=1\linewidth,left]{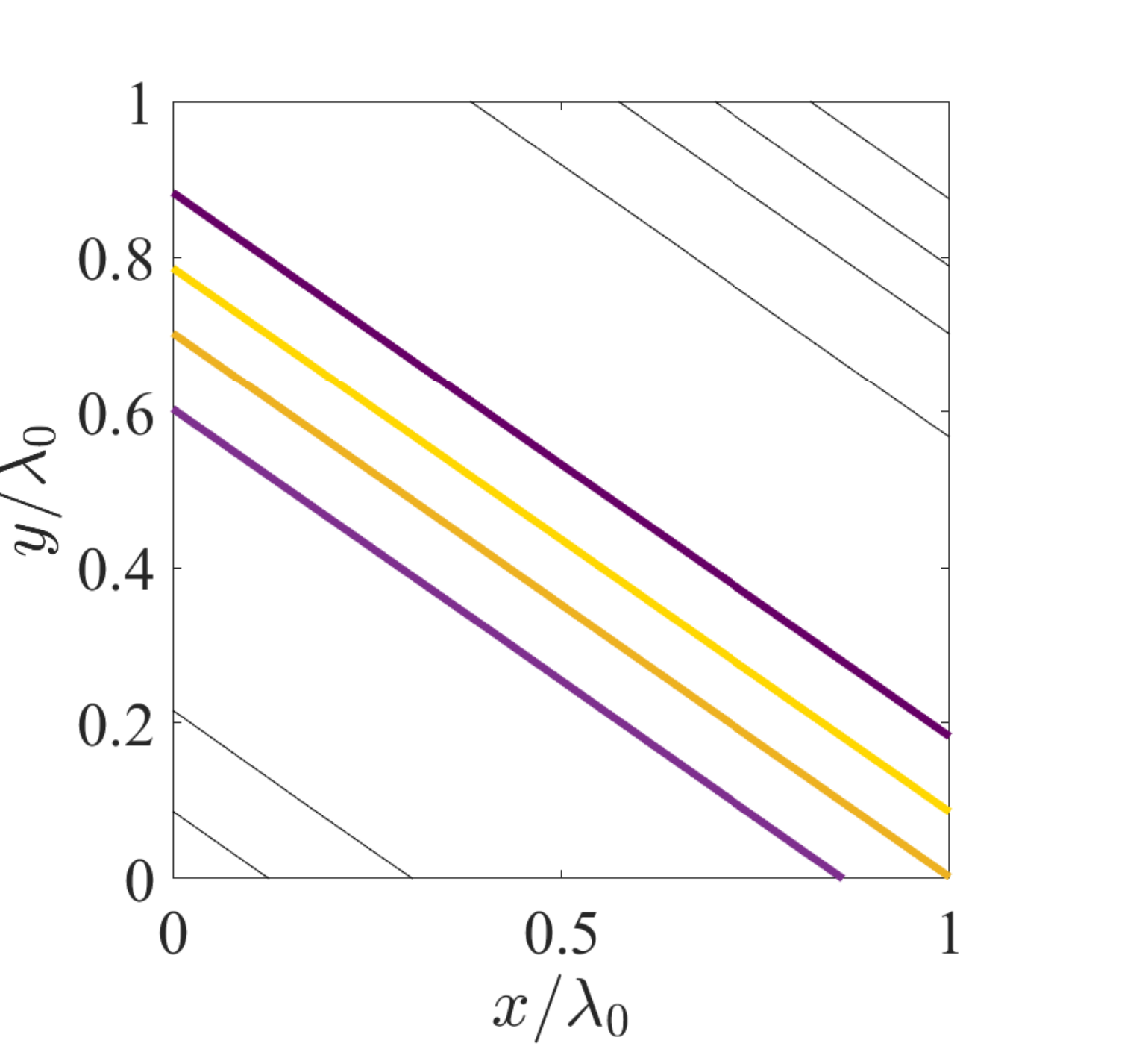}\label{fig:SpecularreflectionB}}
	\endminipage\hfill
	\minipage{0.32\textwidth}
	\subfigure[]{\includegraphics[width=1\linewidth,left]{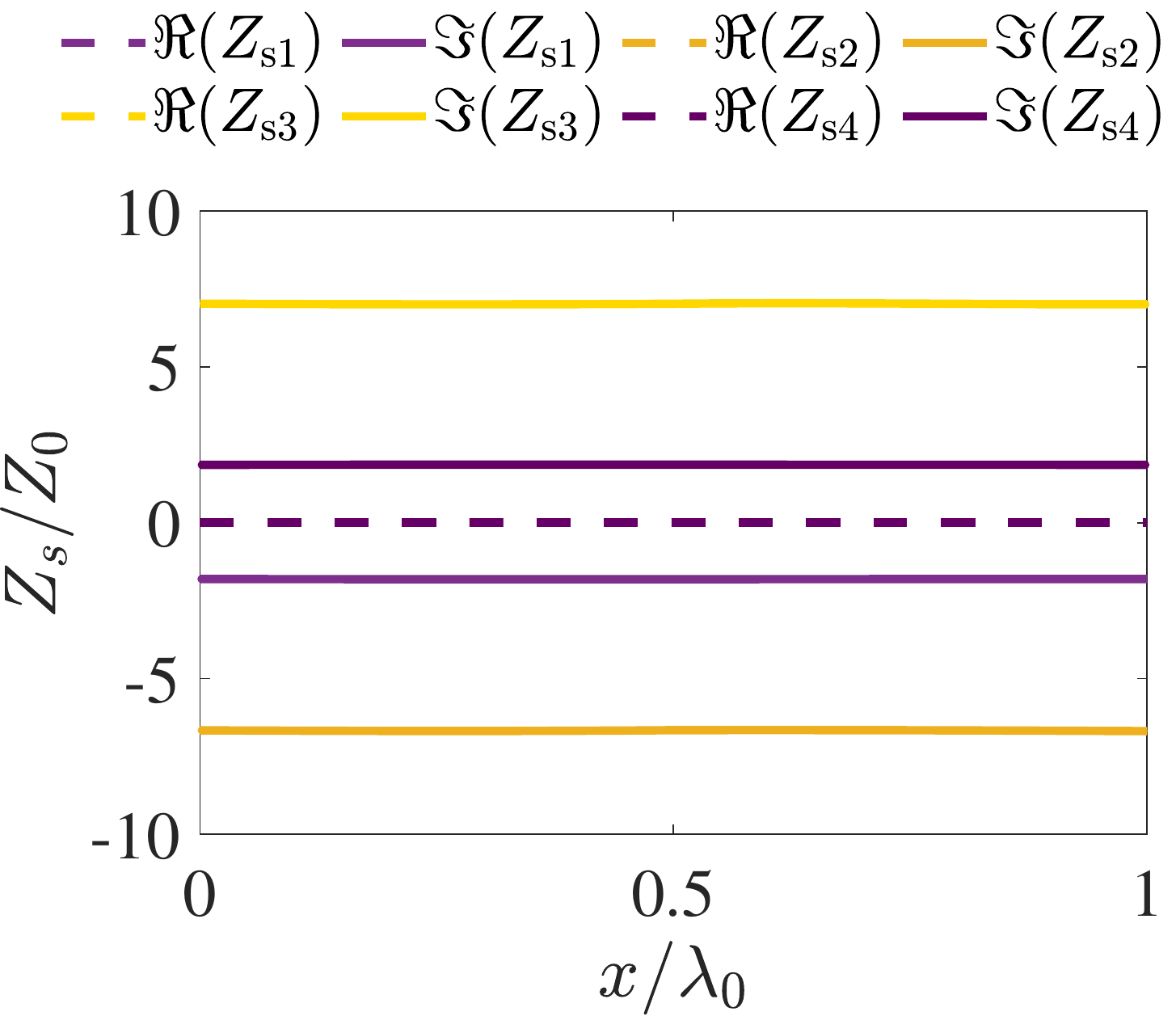}\label{fig:SpecularreflectionC}}
	\endminipage\hfill
	\caption{Results for sets of two plane waves with $\vert p_{\rm i}\vert=\vert p_{\rm r}\vert$, $\theta_{\rm i}=0^\circ$, and $\theta_{\rm r}=70^\circ$. (a) Power flow distribution. (b) Set of surfaces tangential to the intensity vector and (c) the corresponding surface impedance associated with each surface. The four surfaces are denoted by subscripts 1,2,3, and 4.  }
	\label{fig:Specularreflection}
\end{figure*}

\subsection{Retroreflection}

The first case under study is the retroreflection scenario, i.e., we study surfaces capable of redirecting all the energy of an incident plane wave into the opposite direction ($\bm{k}_{\rm r}=-\bm{k}_{\rm i}$), back to the source.  First, we will analyse  acoustic retroreflectors. In order to warranty the power conservation,  the amplitudes of incident and reflected waves should be equal, $\vert p_{\rm i}\vert=\vert p_{\rm r}\vert$. 
By substituting these values in Eqs.~\ref{eq:Ix_2PW}~and~\ref{eq:Iy_2PW}, we can see  that the intensity power-flow vector is zero at all points of space, $I_x(x,y)=I_y(x,y)=0$. It means that in this special case we can design local retroreflectors with any shape, as the power will never cross the reflector boundary.  

Figure~\ref{fig:Retroreflector} shows three different retroreflectors designed for $\theta_{\rm i}=70^\circ$. The first example corresponds to a flat retroreflector where the surface profile is defined as $y_{\rm c}=0$. Figure~\ref{fig:RetroreflectorB} shows the required normalized surface impedance. In the acoustic case, the impedance is normalized with respect to $Z_0=Z_0^{\rm ac}=c_{\rm ac}\rho$. The surface impedance function defines the period of the reflector that reads  $D=\lambda/(2\sin{\theta_{\rm i}})$. Results of numerical simulations of the reflected pressure field are shown in Fig.~\ref{fig:RetroreflectorA} where we can see the reflected plane wave propagating into the opposite direction with respect to that of the incident plane wave.  The second example is a retroreflector with a cosine surface profile $y_{\rm c}=\frac{\lambda}{5}\cos{(2\pi x/D)}$ with the same period as the flat reflector. Figures~\ref{fig:RetroreflectorC} and  \ref{fig:RetroreflectorD} show the reflected field and the normalized surface impedance for this design. We can see that the surface impedance is also purely imaginary confirming the local nature of the design.

The last example is a retroreflector with a cosine modulation where the period is double of that for the flat reflector:  $D'=2D$. The curve which describes this surface modulation profile can be expressed as $y_{\rm c}=\frac{\lambda}{5}\cos{(\pi x/D)}$. The reflected field and the corresponding surface impedance are shown in Figs.~\ref{fig:RetroreflectorE} and \ref{fig:RetroreflectorF}.
It is important to mention that the spatial periodicity of the metasurface can be used as a control parameter for the diffracted modes in the system, allowing us to control reflection for illuminations from other directions.

For design of electromagnetic retroreflectors, even though the physical meaning of the surface impedance is different, we can use the  mathematical analogy with the acoustic scenario.
The correspondence between normalized acoustic and electromagnetic impedances for TE-polarized waves is direct:  $Z_s^{ac}/Z_0^{\rm ac}=Z_s^{TE}/Z_0^{\rm EM}$, with $Z_0^{\rm ac}=c_{\rm ac}\rho$ being the acoustic  wave impedance and  $Z_0^{\rm EM}=c_{\rm EM}\mu$ the electromagnetic wave impedance. Here, $c_{\rm ac}$ and  $c_{\rm EM}$ are the speed of sound and light, respectively. $\rho$ is the mechanical density of medium. However, for TM-polarized waves we obtain, due to the
duality of the problem, that $Z_s^{ac}/Z_0^{\rm ac}=Z_0^{\rm EM}/Z_s^{TM}=Z_0^{\rm EM}Y_s^{TM}$. As we will show, these relations between required impedances for metamirror surfaces will define certain constrains in the design of multiphysics metamirrors.

\subsection{Specular reflection}

Next, we consider reflection into a wave with the same amplitude,  $\vert p_{\rm i}\vert=\vert p_{\rm r}\vert$, but propagating in another direction, with $\bm{k}_{\rm r}\ne\bm{k}_{\rm i}$. In this case, from Eqs.~(\ref{eq:Ix_2PW}) and (\ref{eq:Iy_2PW}) we can see that in general $I_x(x,y)\ne0$ and $I_y(x,y)\ne0$. 
As an example, Fig.~\ref{fig:SpecularreflectionA} shows the power-flow distribution in two interfering plane waves when $\theta_{\rm i}=0^\circ$ and $\theta_{\rm r}=70^\circ$. It is clear from the analysis of the power flow that the surfaces through which there is no power flow are flat surfaces.  
Figure~\ref{fig:SpecularreflectionB} shows a set of surfaces which are  tangential to the intensity vector. The inclination angle  can be written as $\theta_{\rm s}=\arctan\left(\frac{\sin\theta_{\rm i}+\sin \theta_{\rm r}}{-\cos\theta_{\rm i}+\cos \theta_{\rm r}}\right)$. The components of the unit normal vector  $\bm{n}=n_x\hat{\bf x}+n_y\hat{\bf y}$ are defines as $n_x=\frac{1}{\sqrt{2}}(\sin\theta_{\rm i}+\sin \theta_{\rm r})$
and $n_y=\frac{1}{\sqrt{2}}(\cos\theta_{\rm i}-\cos \theta_{\rm r})$. Clearly, these results correspond to trivial specular reflection from these flat surfaces.

The required impedances for each surface are represented in Fig.~\ref{fig:SpecularreflectionC}. We see that the surface impedances are purely imaginary and constant along the surfaces. Depending on the defined surface, the surface impedance takes different values to ensure the desired phase shift between the  incident and reflected waves. We can find positions of surfaces acting as perfect electric conductors (PEC),  where $Z_{\rm s}=0$ which corresponds to surfaces at the planes of zero intensity vector. On the contrary, surfaces with the behavior of perfect magnetic conductor (PMC)  characterized by $Z_{\rm s}=\infty$ are located at the planes of the maximum intensity vector.

\begin{figure*}
\minipage{0.32\textwidth}
	\subfigure[]{\includegraphics[width=1\linewidth,left]{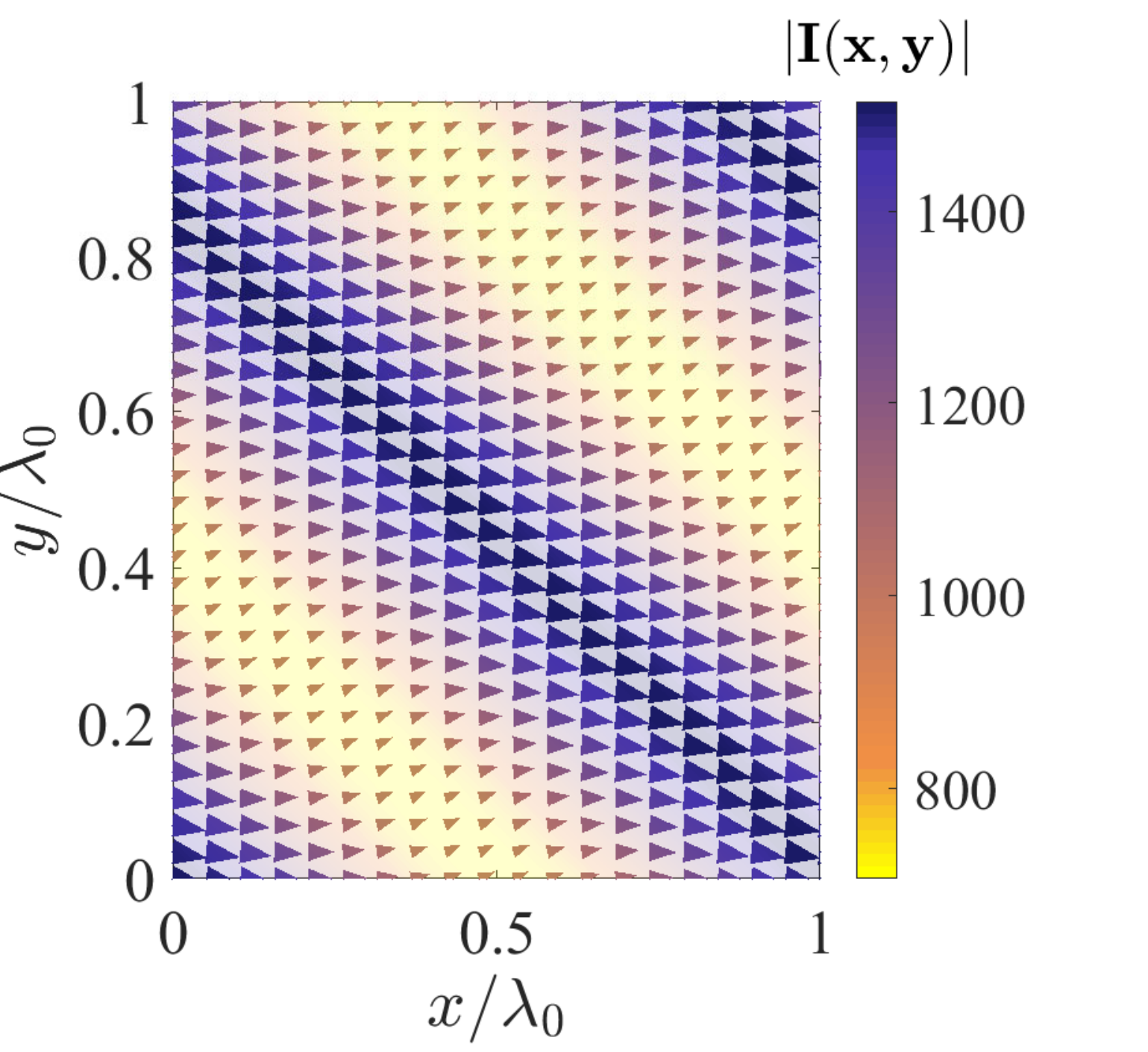}\label{fig:AnomalousReflectorA}}
	\subfigure[]{\includegraphics[width=1\linewidth,left]{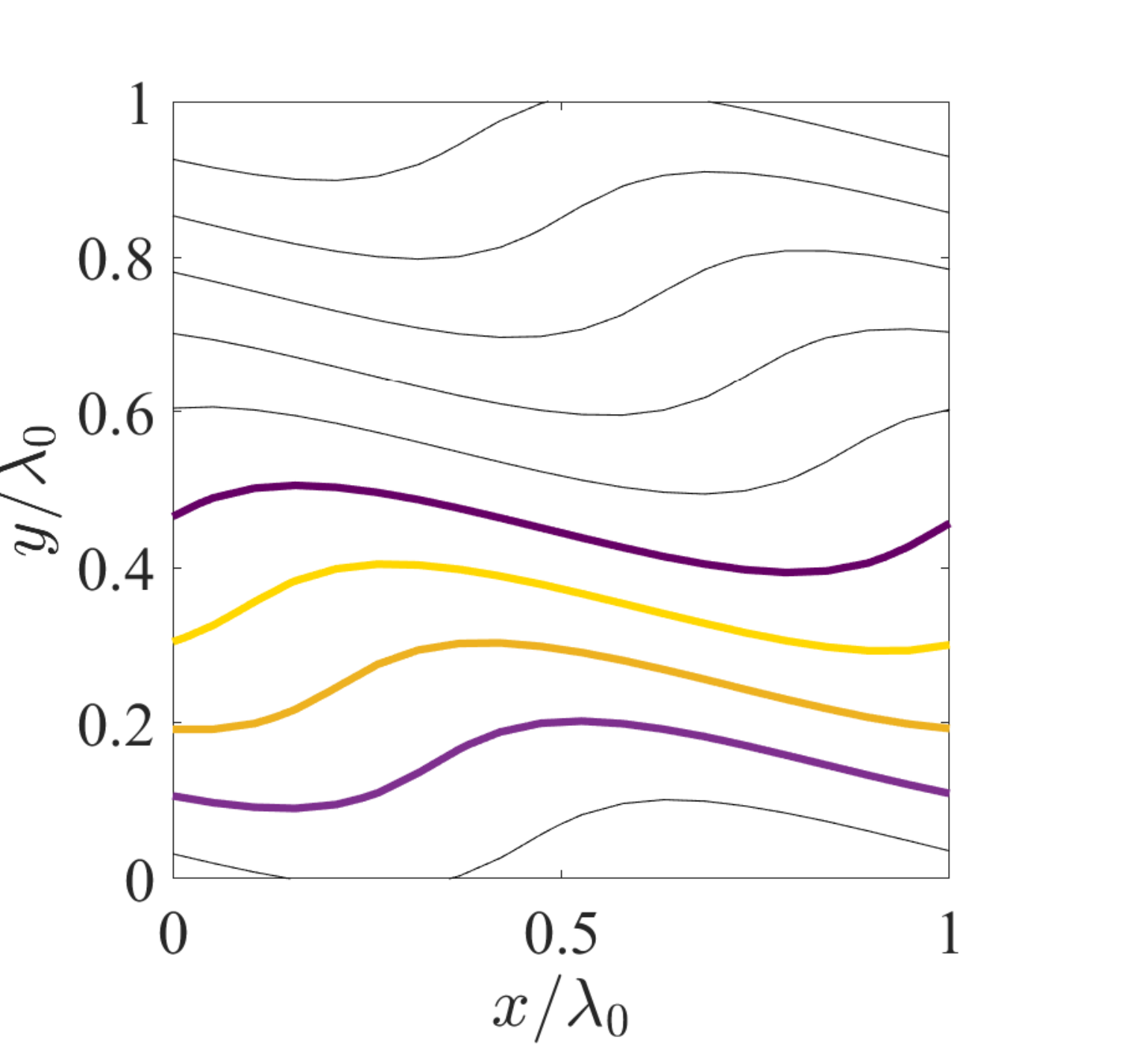}\label{fig:AnomalousReflectorC}}
\endminipage\hfill
	\minipage{0.32\textwidth}
	\subfigure[]{\includegraphics[width=.9\linewidth,left]{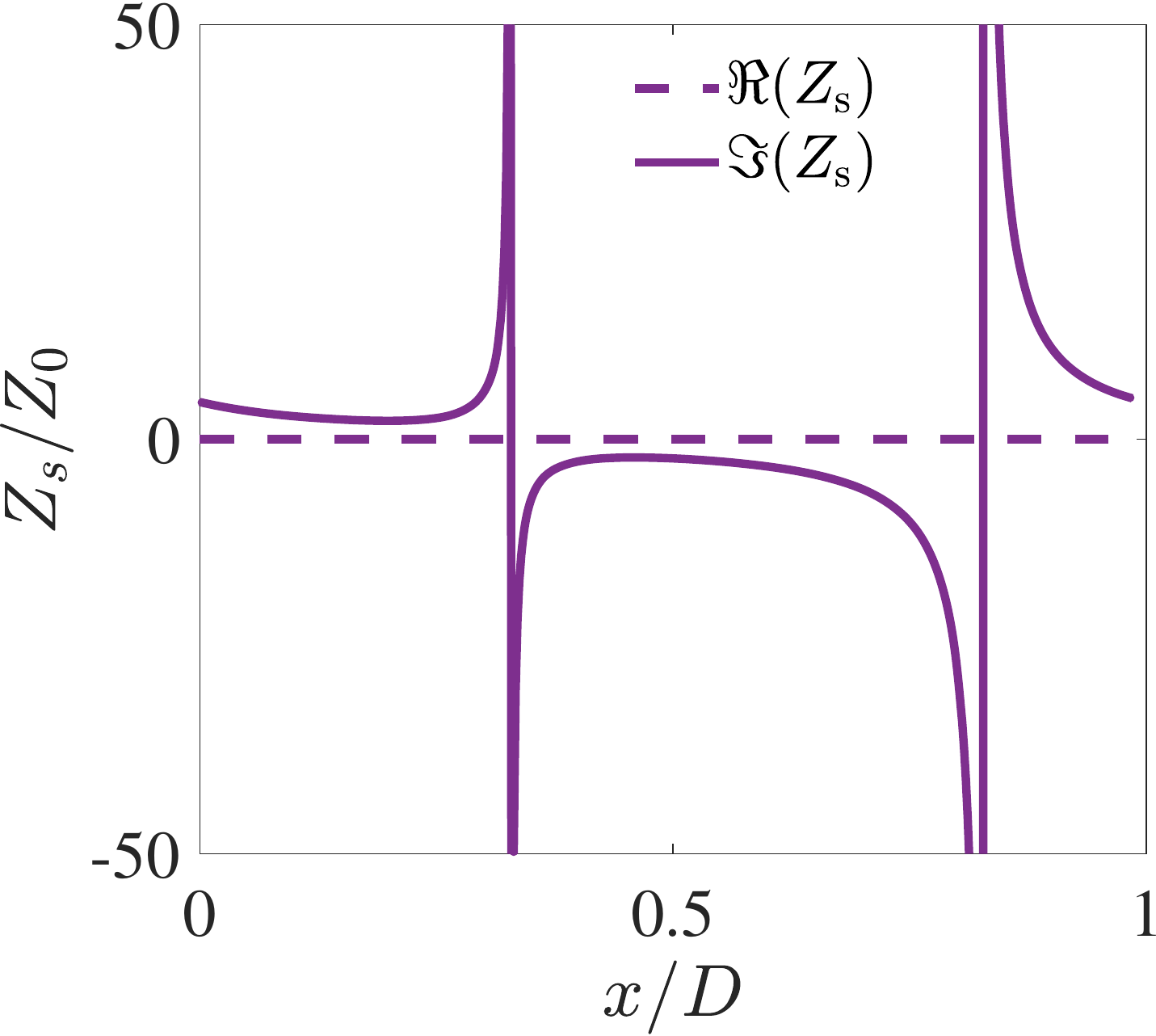}\label{fig:AnomalousReflectorC}}
	\subfigure[]{\includegraphics[width=1\linewidth,left]{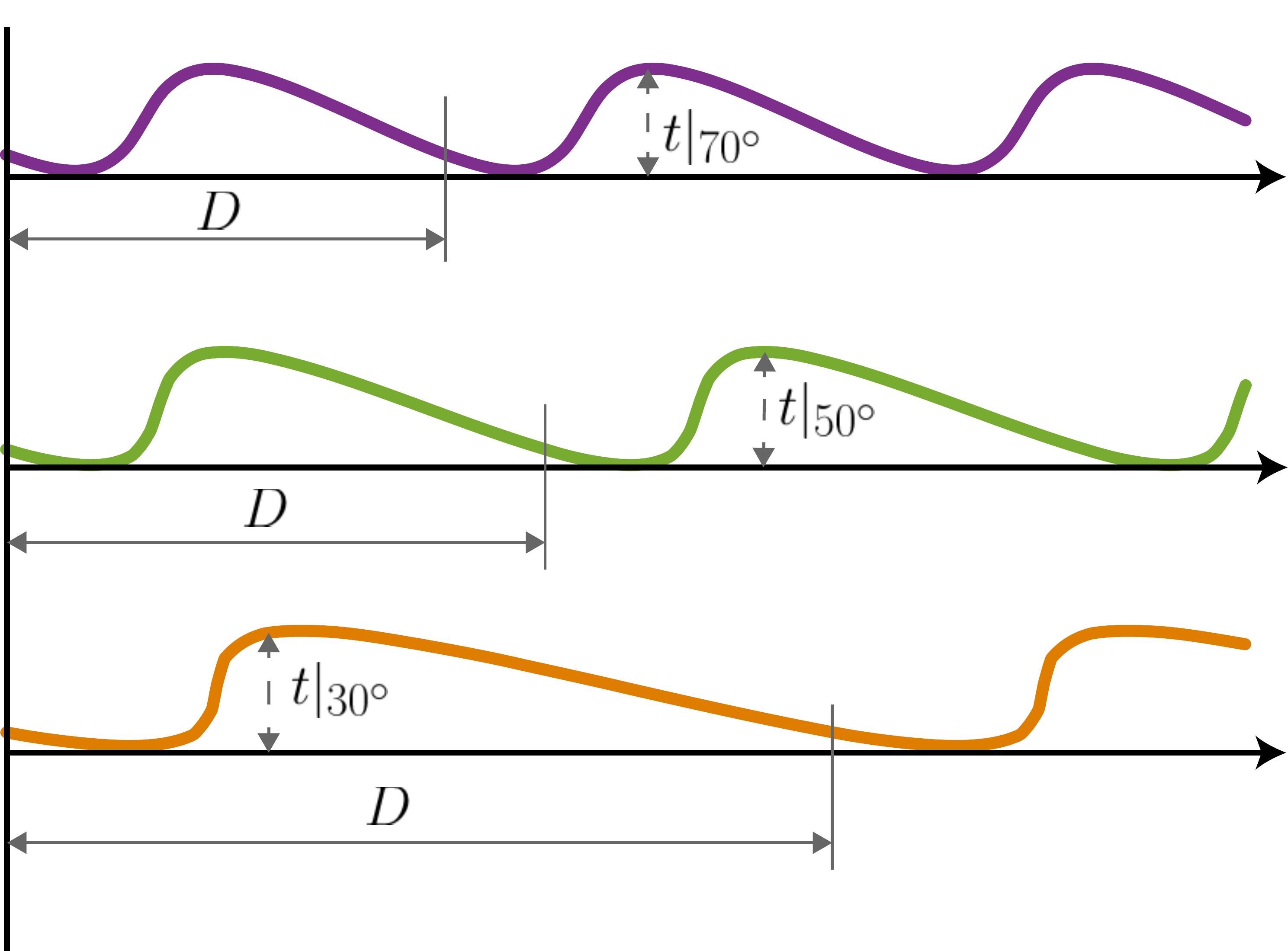}\label{fig:AnomalousReflectorB}}
	\endminipage\hfill
	\minipage{0.32\textwidth}
	\subfigure[]{\includegraphics[width=0.9\linewidth,left]{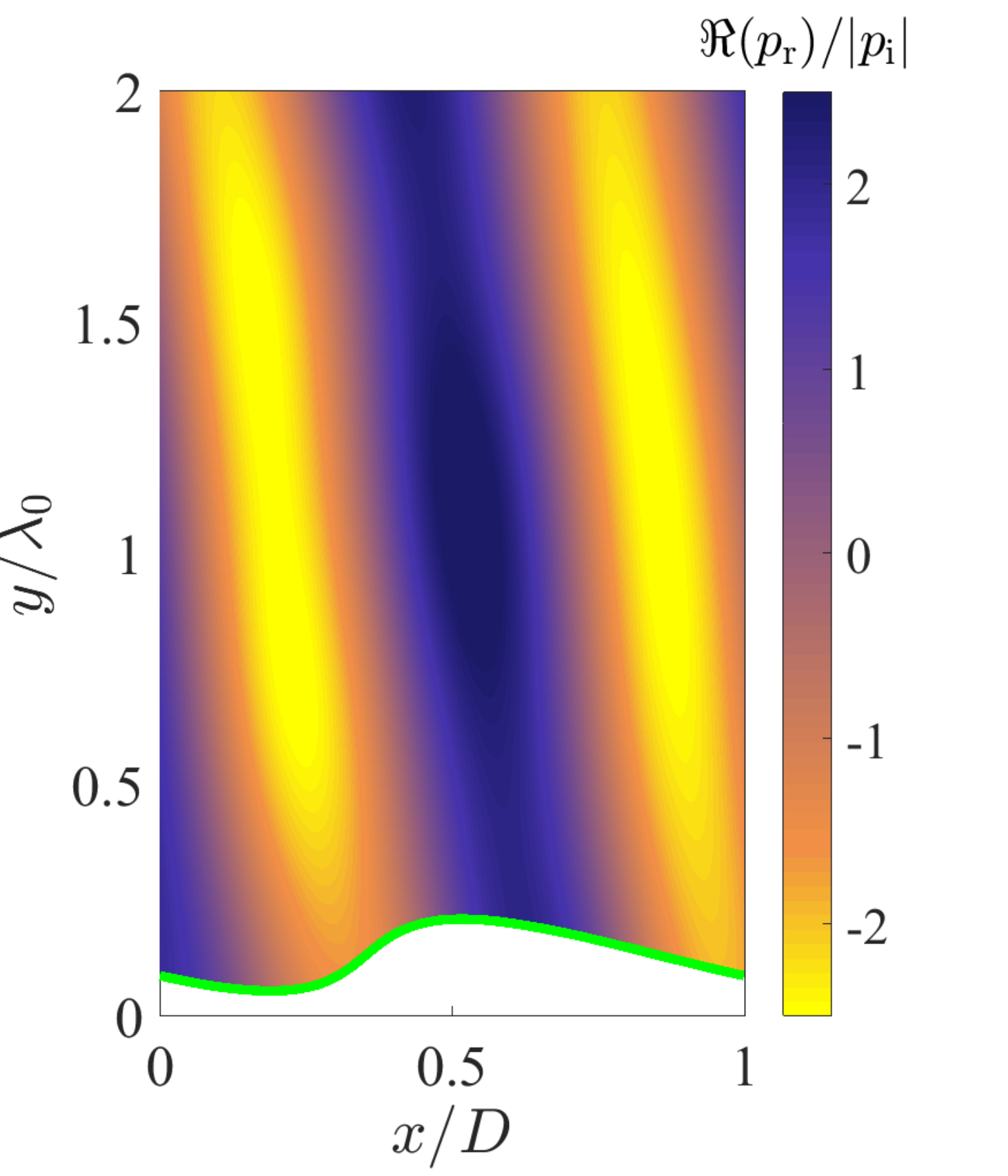}\label{fig:AnomalousReflectorD}}
	\subfigure[]{\includegraphics[width=0.9\linewidth,left]{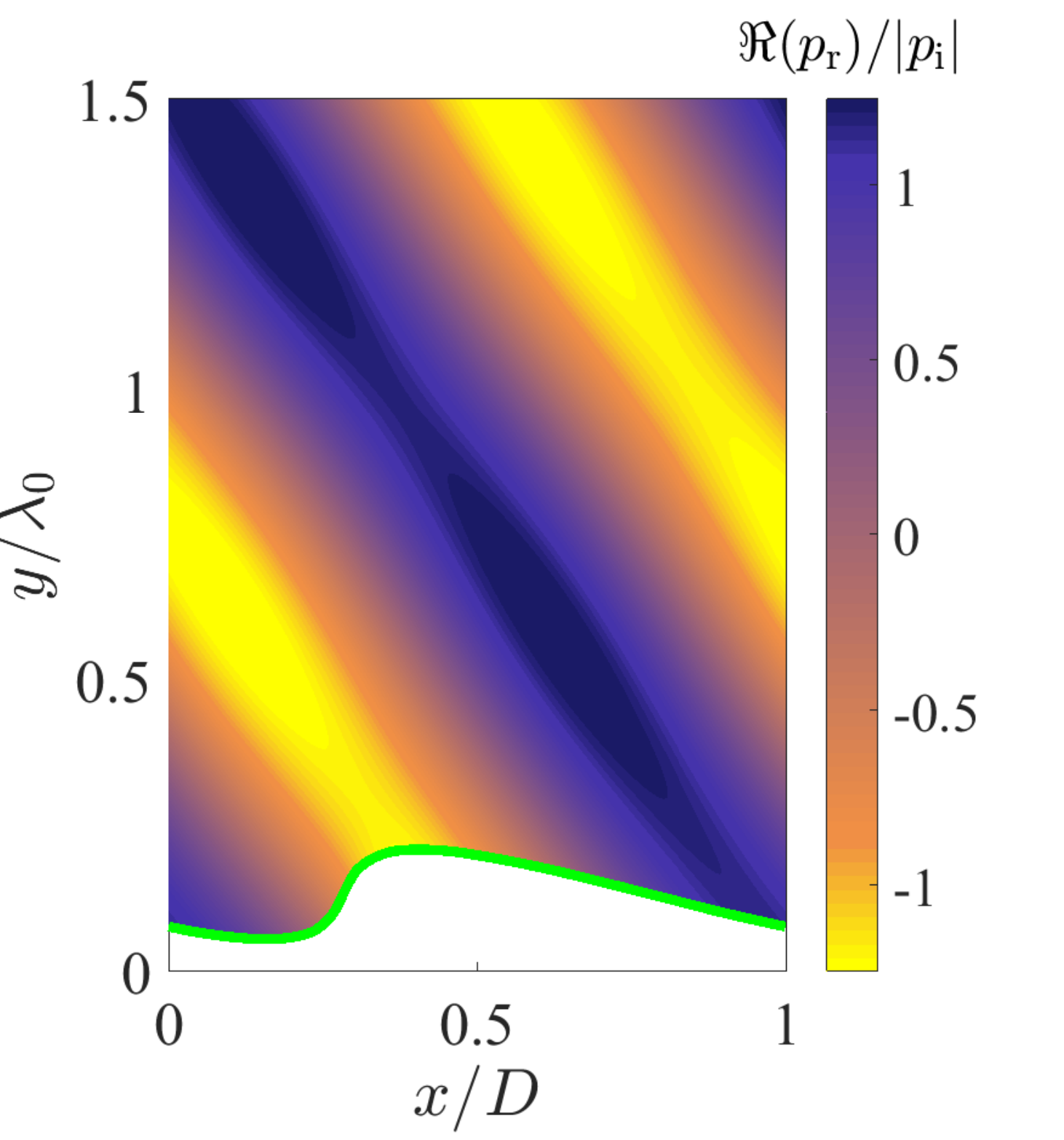}\label{fig:AnomalousReflectorE}}
   \endminipage\hfill	
  
	\caption{ Power flow-conformal metamirror for acoustic  anomalous  reflectors. (a) Power-flow distribution for  $\theta_{\rm i}=0^\circ$ and $\theta_{\rm r}=80^\circ$. (b) Surface profiles for anomalous reflection when $\theta_{\rm r}=80^\circ$. (c) Surface impedance when  $\theta_{\rm i}=0^\circ$ and $\theta_{\rm r}=80^\circ$. (d) Surface profiles for $\theta_{\rm i}=0^\circ$ and $\theta_{\rm r}=30^\circ$, $50^\circ$, and $70^\circ$.   (e) and (f) Numerical simulations for $\theta_{\rm i}=0^\circ$ and $\theta_{\rm r}=80^\circ$ and $\theta_{\rm i}=0^\circ$ and $\theta_{\rm r}=70^\circ$. }
	\label{fig:AnomalousReflector}
\end{figure*}

\subsection{Anomalous reflection}

The third and the most interesting and general case is lossless reflection of a plane wave into an arbitrary direction, where in general both amplitudes and  propagation directions are different: $\vert p_{\rm i}\vert\ne\vert p_{\rm r}\vert$,  ($\bm{k}_{\rm r}\ne\bm{k}_{\rm i}$). This scenario is known as anomalous reflection. For flat metasurfaces laying in the $xz$-plane, it was demonstrated that the relation between the incident and the reflected amplitudes is given by $R=\vert p_{\rm r}\vert/\vert p_{\rm r}\vert=\sqrt{\cos\theta_{\rm i}/\cos\theta_{\rm r}}$. The power flow distribution for the case of $\theta_{\rm i}=0^\circ$ and $\theta_{\rm r}=80^\circ$ is shown in Fig.~\ref{fig:AnomalousReflectorA}.

As it was proposed in \cite{Diaz_Power_2019}, instead of trying to  synthesize flat surfaces with the required complex-valued impedances, we can create the desired field distribution shaping an impenetrable lossless reflector along surfaces which are tangential to the intensity vector. These surfaces  can be found using  the scalar function
\begin{equation}
	g(x,y)=Ay+B\sin(\Delta\bm{k}\cdot\bm{r}+\Delta\phi)+C, \end{equation}
where $A=I_0\vert p_{\rm i}\vert^2k_0(\sin\theta_{\rm i}+\cos\theta_{\rm i}\tan\theta_{\rm r})$, $B=I_0\vert p_{\rm i}\vert\sqrt{\cos\theta_{\rm i}/\cos\theta_{\rm r}}(\cos\theta_{\rm i}-\cos\theta_{\rm r})/(\sin\theta_{\rm r}-\sin\theta_{\rm i})$, and $C$ is an arbitrary constant. This scalar function satisfies $\nabla g(x,y)=\bm{N}(x,y)$, with vector  $\bm{N}(x,y)=-I_y(x,y)\hat{\bf x}+I_x(x,y)\hat{\bf y}$ being  orthogonal  to the intensity vector. From the properties of the gradient, the surfaces tangential to the power flow directions can be calculated as the level curves of the function $g(x,y)$ and can be expressed as $y_{\rm c}=f(x)$.

Probably the most studied case  in the literature of anomalous reflection in  a normally incident plane wave ($\theta_{\rm i}=0$) that is redirected into an arbitrary direction ($\theta_{\rm r}\ne 0$). In this case, the surface profile and the surface impedance are different for each specific angle of reflection.
Figure~\ref{fig:AnomalousReflectorB}  shows the shapes of the perfectly anomalously reflecting surfaces for $\theta_{\rm r}=30^\circ,50^\circ,$ and $70^\circ$. 
As can be expected, the period of the surface modulation changes with the reflected angle according to $D=\lambda/(\sin \theta_{\rm r}-\sin \theta_{\rm i})$. The amplitude of the modulation, $h$, keeps almost constant for different incident angles ($h\vert_{30^\circ}=0.1566\lambda$, $h\vert_{50^\circ}=0.1538\lambda$, and $h\vert_{70^\circ}=0.1386\lambda$), and is always subwavelength.  The required surface impedances for these particular surfaces are plotted in Fig.~\ref{fig:RetroreflectorD}. Two numerically simulated examples are presented in Figs.~\ref{fig:RetroreflectorC} and ~\ref{fig:RetroreflectorD}.

\section {Multiphysics implementation}

Here, we will show that using the concept of power flow-conformal metamirrors and the analogy between acoustic and electromagnetic impedances it is possible to create metasurfaces which operate as retro- and most general anomalous reflectors simultaneously for both  acoustic and electromagnetic waves. 

As it was explained above, the  metamirrors functionalities are  defined by the surface profile and the surface impedance. 
First, if we want to control both  acoustic and electromagnetic waves, we need to ensure that the surface profiles required for both operations are compatible. In the previous section, we saw  that some scenarios, such as anomalous reflection, require specific  surface profiles. However, in other cases there is no such restriction and one can use different surface profiles for the same functionality, like for realizing retroreflectors. With this on mind, we distinguish two different design approaches: (i) Platforms that implement the same functionality for acoustic and electromagnetic waves and (ii) Platforms with different functionalities for acoustic and electromagnetic waves. 

\begin{figure}[]
	\includegraphics[width=1\columnwidth,left]{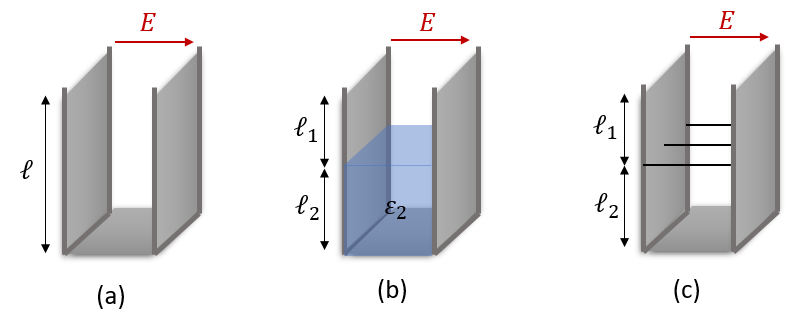}
	\caption{ Multiphysics meta-atoms. (a) Empty closed-end groves. (b) Partially filled closed-end groves. (c) Closed-end groves with a metallic grid. }
	\label{fig:Metaatom}
\end{figure}

The second consideration is the design of the meta-atoms for implementing the corresponding surface impedance.
Power flow-conformal metamirrors are described by locally defined acoustic of electromagnetic surface impedances, which allows  implementations  using conventional phase shifters of different types. 
The main challenge in the multiphysics realizations is to find such meta-atom configurations which will provide desired phase shifts both for sound and electromagnetic waves. We have identified three different structures which can be used for achieving this goal. The choice between them will be made based on specific conditions in each design. The schematic representations of the proposed three topologies are shown in Fig.~\ref{fig:Metaatom}. 

The first meta-atom a closed-end grove [see Fig.~\ref{fig:Metaatom}(a)]. Since here we consider surfaces which are modulated only along one direction, we assume that the groves are infinitely long and uniform. In general, two-dimensional arrays or tubes can be used. If the thickness of the walls is much  smaller than the width of the groves, we can neglect the effect of wall thickness on the surface impedance. The acoustic  surface impedance $Z_{\rm s}^{\rm ac}=-jZ_0^{\rm ac}\cot(k_{\rm ac}\ell)$ is controlled by the depth of the groves $\ell$  and the acoustic impedance $Z_0^{\rm ac}=c_{\rm ac}\rho$ of air filling the tube. Here,  $k_{\rm ac}=2\pi f_{\rm ac}/c_{\rm ac}$ is the acoustic wavenumber. For  TE-polarized electromagnetic waves  (electric field parallel to the grove), in the microwave range an array of groves reflects as a practically perfectly conducting surface (PEC) if the walls are made of metal and the width of the groves is deeply subwavelength.  For TM-polarized electromagnetic waves, the response of the meta-atom is defined  by the surface impedance $Z_{\rm s}^{\rm EM}=jZ_0^{\rm EM}\tan(k_{\rm EM}\ell)$, with $Z_0^{\rm EM}=\sqrt{\mu_0/\varepsilon_0}$ and $k_{\rm EM}=2\pi f_{\rm EM}/c_{\rm EM}$ being the electromagnetic wave impedance and the wavenumber in air, respectively.

The second proposed topology is the same closed-end grove but partially  filled with a dielectric material with permitivitty $\varepsilon_2$. The dielectric material should be chosen to behave as a hard boundary for acoustic waves, introducing  strong impedance contrast with air for acoustic waves. In practice, one can consider solid plastics or water. As we can see in Fig.~\ref{fig:Metaatom}(b), the filling with length $\ell_2$  is placed at the bottom of the grove. The length of the air layer is denoted as $\ell_1$. The acoustic response of this meta-atom is defined by $Z_{\rm s}^{\rm ac}=-jZ_0^{\rm ac}\cot(k_{\rm ac}\ell_1)$.  In contrast, the electromagnetic response for TM-polarized waves is given  by the following surface impedance:  
\begin{equation}
Z_{\rm s}^{\rm EM}=jZ_0^{\rm EM}\frac{\tan(k_{\rm EM}^{(2)}\ell_2)+\tan(k_{\rm EM}\ell_1)}{1-\tan(k_{\rm EM}^{(2)}\ell_2)\tan(k_{\rm EM}\ell_1)}.\label{eq7}
\end{equation}
Thus, the use of this structure allows us to independently tune both acoustic and electromagnetic impedances to the required values.

The third proposed topology is shown in Fig.~\ref{fig:Metaatom}(c). In this case, we use a grove  of the  total depth $\ell=\ell_1+\ell_2$ with an array of  metal wires  oriented in the same direction as the electric field of TM-polarized waves and positioned at  distance $\ell_2$ from the bottom of the grove. The array of metal wires has a period $d_{\rm w}$ small enough for the array to  act as a nearly perfect electric wall for electromagnetic waves. The reflection of electromagnetic waves is very strong when the period of the array of wires is much smaller than the wavelength of electromagnetic waves, while the diameter of the wires is still very small compared with the period \cite{modeboo}. This property allows us to realize a wire array with is nearly perfect reflector for electromagnetic waves which is nearly perfectly transparent for sound.  The electromagnetic response of this meta-atoms is characterized by the surface impedance $Z_{\rm s}^{\rm EM}=jZ_0^{\rm EM}\tan(k_{\rm EM}\ell_1)$ while the acoustic response is defined  by $Z_{\rm s}^{\rm ac}=-jZ_0^{\rm ac}\cot(k_{\rm ac}\ell)$. Again we see that there is a possibility to adjust both impedances independently.

Using these meta-atoms we can systematically design various devices  for simultaneous control of electromagnetic and acoustic waves.


\subsection{The same  functionality in one platform}

Here we explore possibilities to implement metamirrors with the same functionality for both electromagnetic and acoustic waves. In particular, we will use the proposed meta-atoms  to implement dual-physics retroreflectors and anomalous reflectors.

\begin{figure}[t]
\minipage{0.47\columnwidth}
	\subfigure[]{\includegraphics[width=1\linewidth,left]{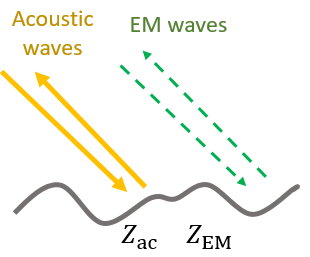}\label{fig:Fig6A}}
	\subfigure[]{\includegraphics[width=1\linewidth,left]{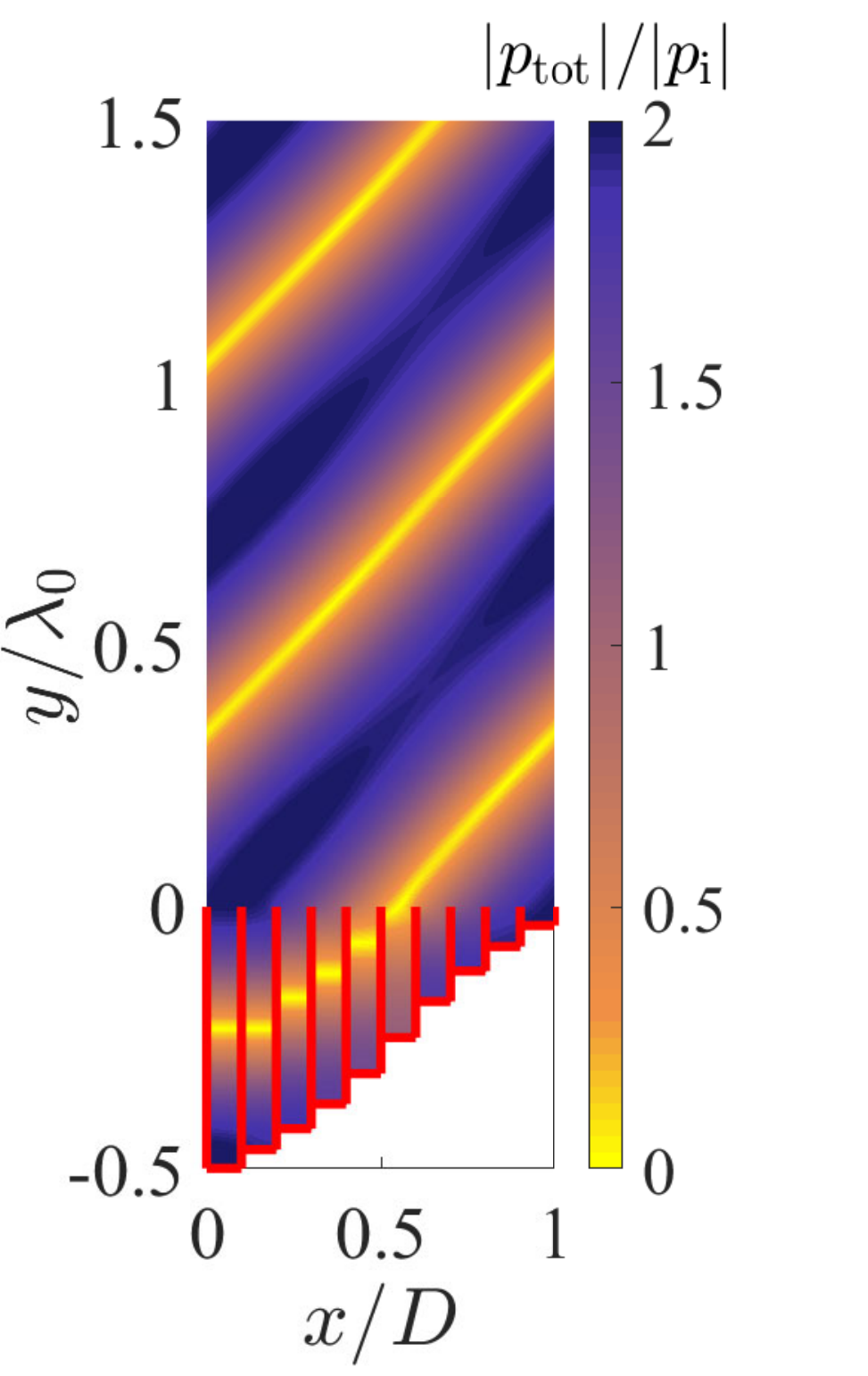}\label{fig:Fig6B}}
\endminipage\hfill\minipage{0.47\columnwidth}
	\subfigure[]{\includegraphics[width=1\linewidth,left]{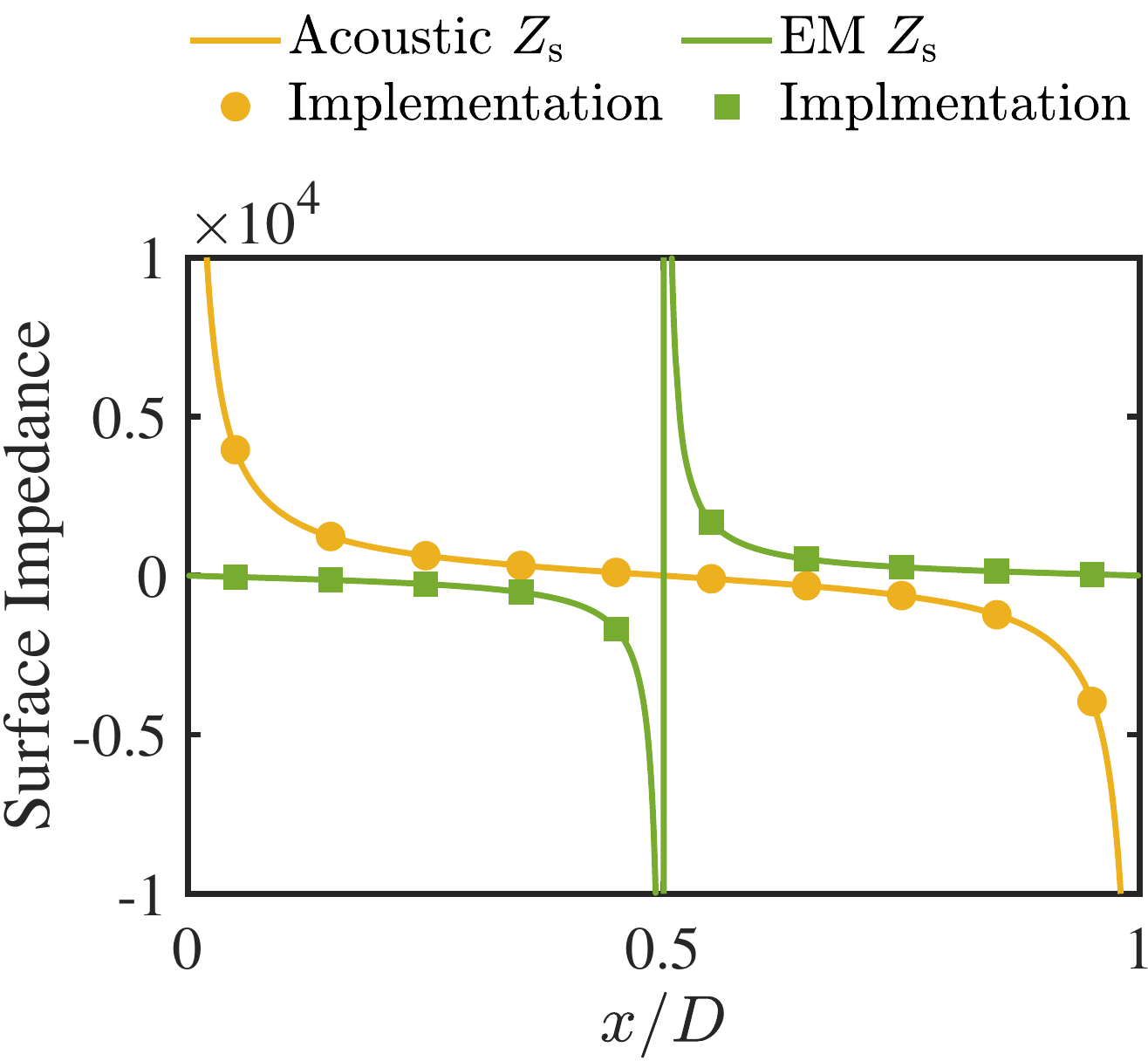}\label{fig:Fig6C}}
	\subfigure[]{\includegraphics[width=1\linewidth,left]{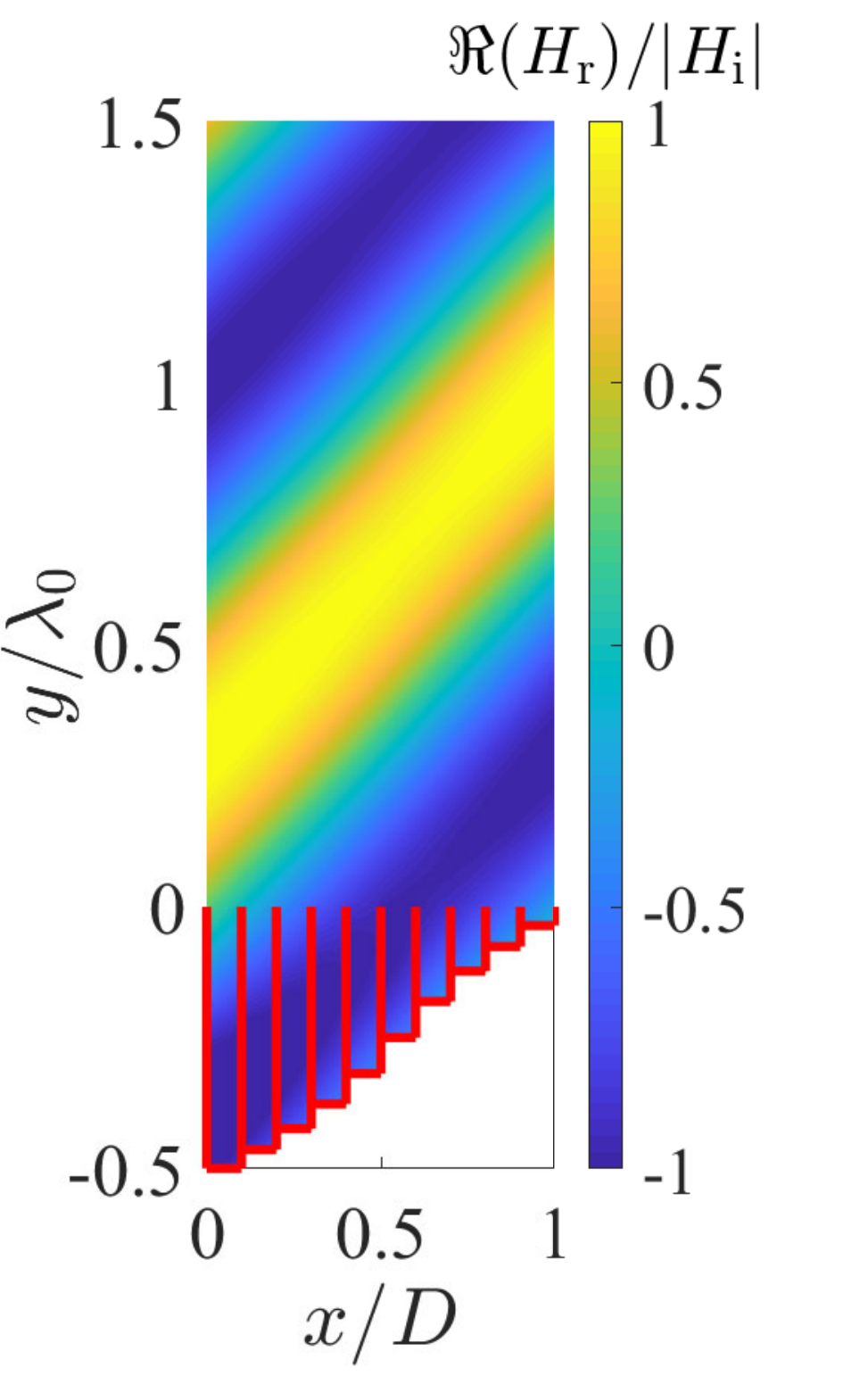}\label{fig:Fig6D}}
\endminipage\hfill
	\caption{Dual-physics retroreflector. (a) Schematic representation of the metamirror. (b) Total pressure field calculated with a numerical simulation for acoustic waves. Red lines denote the hard boundary conditions used for simulating the groves. (c) Surface impedance for electromagnetic (EM) and acoustic metamirror.  (d) Reflected magnetic field calculated with a numerical simulation for EM waves. Red lines denote the PEC boundary conditions used for simulating the groves. }
	\label{fig:Case1}
\end{figure}

\textit{\underline{Dual-physics retroreflector}:} Retroreflectors are able to send all the incident energy back into the same direction from where the incident wave is coming [see Fig.~\ref{fig:Case1}(a)]. Following the design methodology explained above, we will design a retroreflector working for acoustic and $TM$-polarized electromagnetic waves. 
The first step in the design of the multidisciplinary retroreflectors is to choose an appropriate reflector profile. In this work, for simplicity we use the simplest surface profile: a flat surface located in the $xz$-plane. As shown above, this is possible because the desired field structure is a purely standing wave without any power flow across any surface. 

Once the retroreflecting surface position is defined, the periodicity is fixed by the operation frequencies, both in acoustics, $f_{\rm ac}$, and in electromagnetism, $f_{\rm EM}$. We start from considering the simplest case when the frequencies are such that $\lambda_{\rm ac}=\lambda_{\rm EM}$, i.e., the periodicity for both electromagnetic and acoustic scenarios is the same ($D_{\rm ac}=D_{\rm EM}$). To satisfy this condition, the frequencies should be related as $f_{\rm EM}=f_{\rm ac}c_{\rm EM}/c_{\rm ac}$. As an example, we assume that the background medium is air characterized by $c_{\rm EM}=3\cdot10^8 $~m/s and $c_{\rm ac}=343 $~m/s (the speed of sound in the background media has been calculated using a reference temperature of $T=20^\circ {\rm C}$). 

The last step in the design is to implement the required surface impedances. For flat retroreflectors, the required surface impedances for electromagnetic and acoustic waves can be written as 
\begin{eqnarray}
Z_s^{\rm ac}(x)=j\frac{Z_0^{\rm ac}}{\cos{\theta_{\rm i}}} \cot{(k_{\rm ac}\sin{\theta_{\rm i}} x)}   \label{eq:CASE1_1} \\
Z_s^{\rm EM}(x)=-jZ_0^{\rm EM}\cos{\theta_{\rm i}} \tan{(k_{\rm ac}\sin{\theta_{\rm i}} x)}\label{eq:CASE1_2}
\end{eqnarray}
The surface impedance values when $\theta_{\rm i}=70^\circ$ are represented in Fig~\ref{fig:Fig6C}.  These surface impedance profiles can be  implemented  using empty  closed-end
groves as meta-atoms. The depths of the groves are 50~mm, 46.4~mm, 42.4~mm, 37.7~mm, 31.9~mm, 25~mm, 18.1~mm, 12.3~mm, 7.6~mm, and 3.6~mm. Figures~\ref{fig:Case1}(b) and (d) show the results of  numerical simulation of a multidisciplinary retroreflector for  $f_{\rm ac}=3430$~Hz and $f_{\rm EM}=3$~GHz.  From the distribution of the total acoustic field represented in Fig.~\ref{fig:Fig6B} we can see how a standing waves is generated as a consequence of  interference between incident and reflected waves. Figure~\ref{fig:Fig6D} shows the reflected field in the electromagnetic scenario where we can see that all the energy is sent back along the propagation axis of the incident wave.
 

If the acoustic and electromagnetic frequencies are arbitrary, and the corresponding wavelengths are not equal, the required periodicity of the metasurface for  each scenario will be different, $D_{\rm ac}\neq D_{\rm EM}$. In this case, we select the period of the structure to be equal to the  least common multiple of the two periods. It is also important to notice in this general case   the relation between the acoustic and electromagnetic surface impedances is not described by Eqs.~(\ref{eq:CASE1_1}) and (\ref{eq:CASE1_2}). Thus, implementation with simple close-ended groves is not possible. In this case, we should use meta-atoms consisting of partially  filled  closed-end  groves or closed-end groves with a metallic grid, which allow realizations of the required impedance profiles for both fields. Two examples on how to use these meta-atoms are given in Section~III.B

\begin{figure}[t]
\minipage{0.47\columnwidth}
	\subfigure[]{\includegraphics[width=1\linewidth,left]{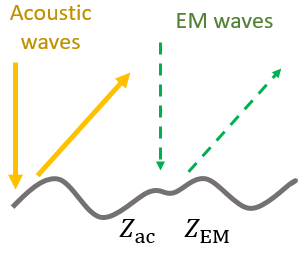}\label{fig:AnomalousReflectorA}}
	\subfigure[]{\includegraphics[width=1\linewidth,left]{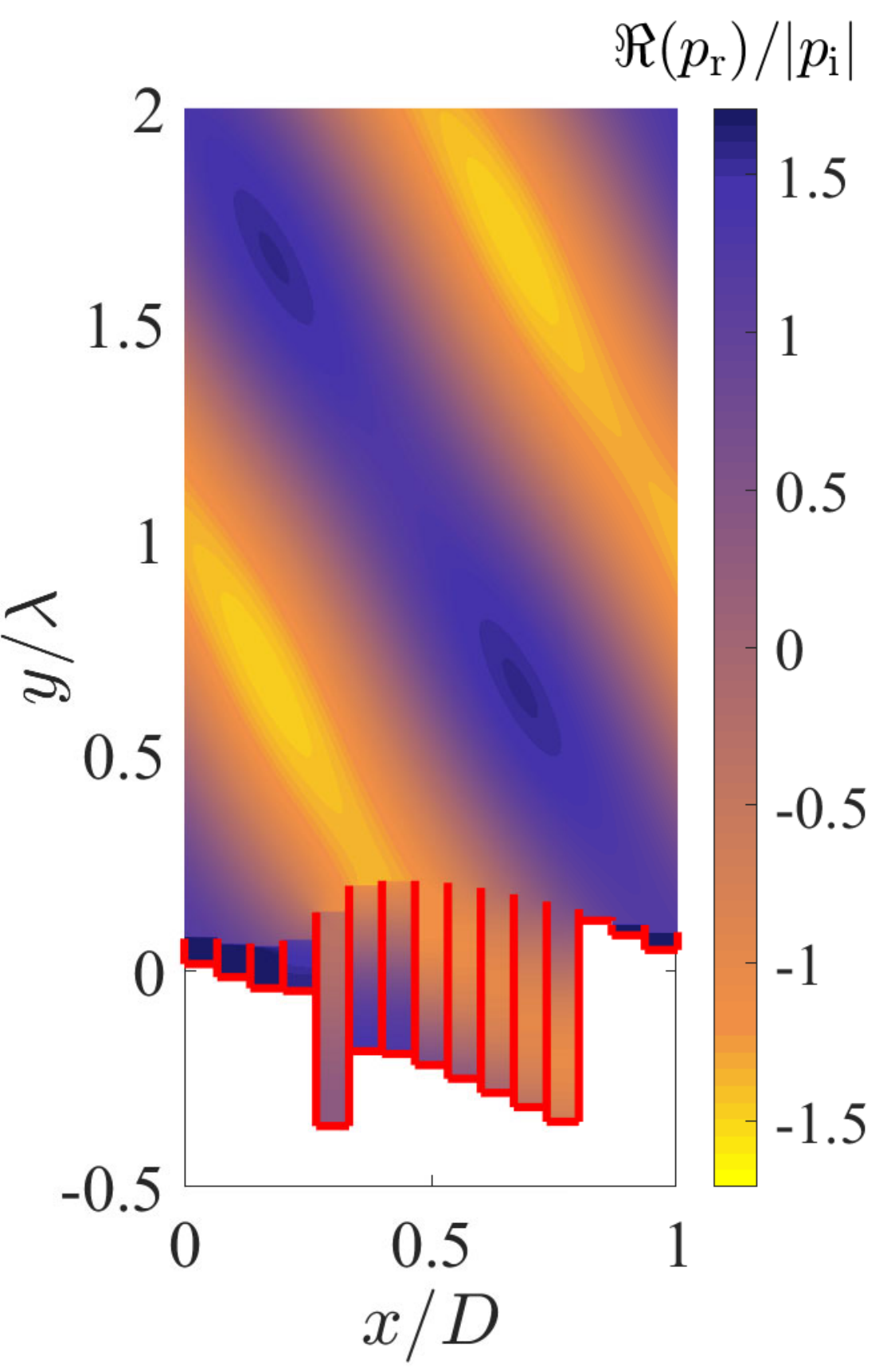}\label{fig:AnomalousReflectorA}}
\endminipage\hfill\minipage{0.47\columnwidth}
	\subfigure[]{\includegraphics[width=1\linewidth,left]{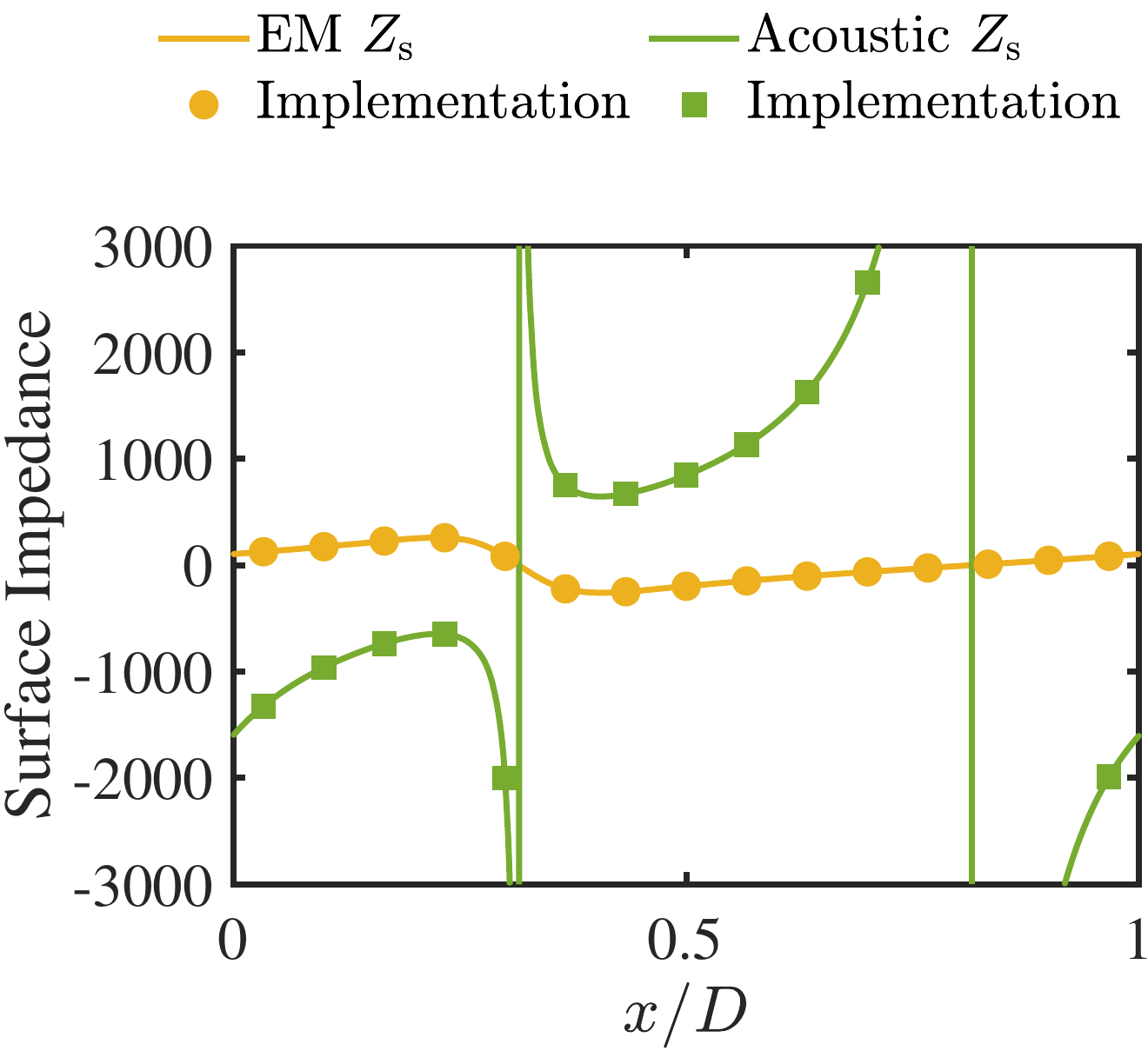}\label{fig:AnomalousReflectorC}}
	\subfigure[]{\includegraphics[width=1\linewidth,left]{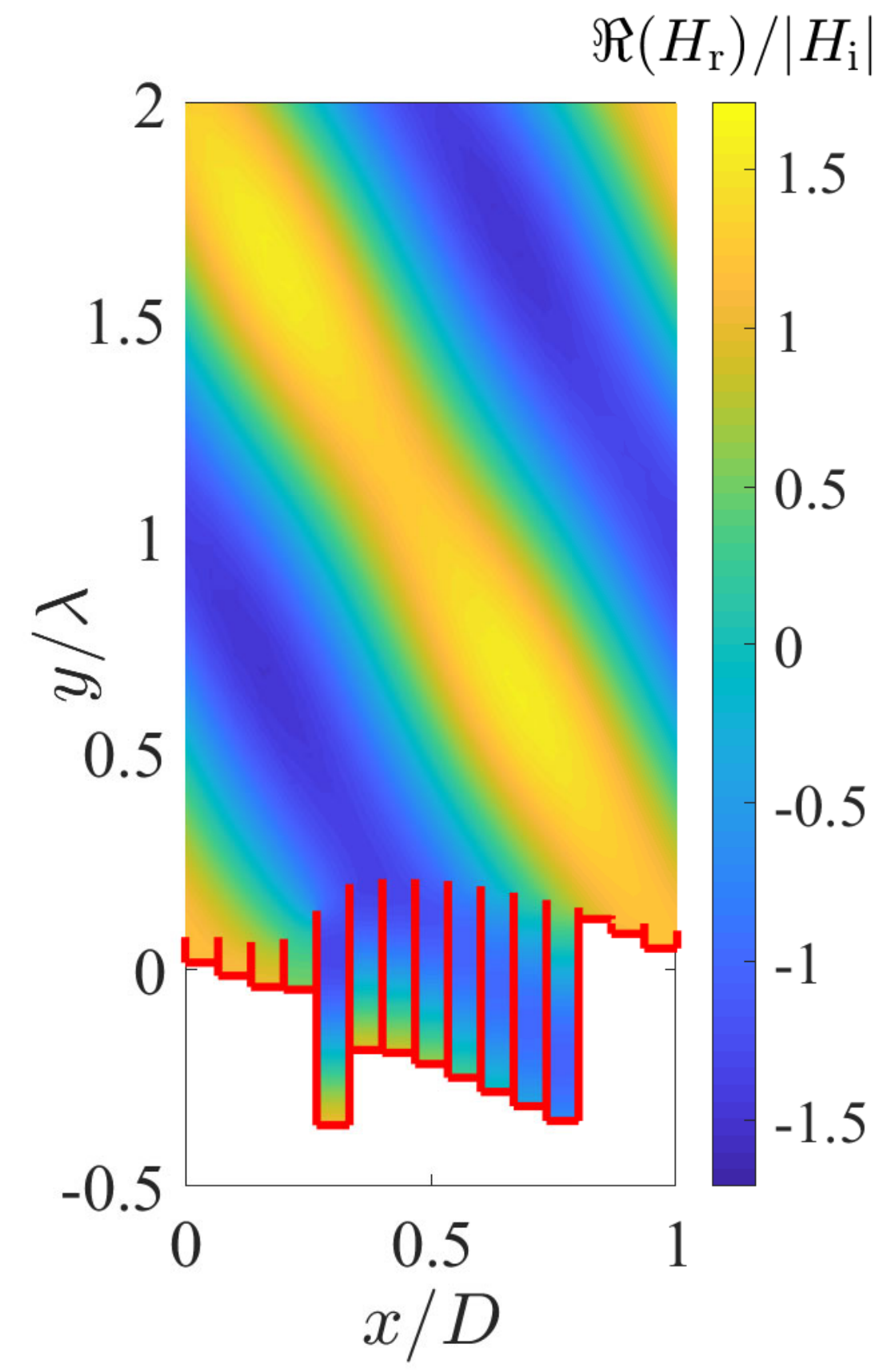}\label{fig:AnomalousReflectorA}}
\endminipage\hfill
	\caption{Dual-physics anomalous reflector. (a) Schematic representation of the metamirror. (b) Reflected pressure field calculated with a numerical simulation for acoustic waves. Red lines denote the hard boundary conditions used for simulating the groves. (c) Surface impedances for electromagnetic (EM) and acoustic metamirror.  (d) Reflected magnetic field calculated with a numerical simulation for EM waves. Red lines denote the PEC boundary conditions used for simulating the grove walls. }
	\label{fig:Case2}
\end{figure}

\textit{\underline{Dual-physics anomalous reflector}:} The second example is an anomalous reflector for acoustic and electromagnetic waves. The schematic representation of the proposed metamirror is shown in Fig.~\ref{fig:Case2}(a). This device is able to send impinging acoustic and electromagnetic plane waves coming from a certain direction into an arbitrary direction. As it was explained in Section~II, in this case there is a  periodic power-flow distribution in the $xy$-plane. Following the design procedure for power flow-conformal metamirrors, the first step is to define an appropriate surface profile for the desired incident angle $\theta_{\rm i}$ and the desired reflection angle $\theta_{\rm r}$, ensuring that the power flow is always tangential to the reflecting surface. 
The second step in the design is to implement the required surface impedances for both acoustic and electromagnetic waves. In this case, we will follow a similar approach as in the previous example and exploit the fact that the normalized $Z_{\rm ac}$ is analogous to normalized $Y_{\rm TM}$. As we have already demonstrated, using empty metallic  close-ended groves we can realize a metasurface which the desired acoustical and electromagnetic properties. In particular, for an anomalous reflector with $\theta_{\rm i}=0^\circ$ and $\theta_{\rm r}=70^\circ$,  the needed depths   of the groves are 5.9~mm, 7.8~mm, 10~mm, 11.7~mm, 49.5~mm, 38.3~mm, 40.1~mm, 42.2~mm, 44.2~mm, 46~mm, 47.6~mm, 49.2~mm, 0.8~mm, 2.4~mm, and 4.1~mm (considering the frequencies   $f_{\rm EM}=3$~GHz and  $f_{\rm ac}=3430$~Hz as an example). Figures~\ref{fig:Case2}(b) and (d) show the scattered fields produced by the structure at the operation frequency for both acoustical and electromagnetic scenarios. 


\subsection{Different functionalities in the same platform}

Previous  examples have shown a possibility to exploit the local behavior of power flow-conformal metasurfaces and the direct analogy between acoustic and electromagnetic waves to implement multidisciplinary meta-devices that produce the same response for acoustic and electromagnetic waves. Next, we propose meta-devices realizing different transformations for acoustic an electromagnetic waves. As particular examples, we will study a retroreflector with different angles for acoustic and electromagnetic waves and an anomalous reflector for acoustic waves that behaves as a retroreflector for electromagnetic waves. 

\begin{figure}[b]
\minipage{0.47\columnwidth}
	\subfigure[]{\includegraphics[width=1\linewidth,left]{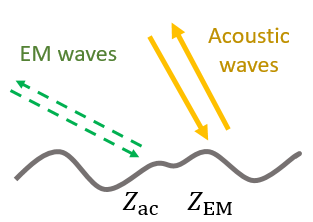}\label{fig:AnomalousReflectorA}}
	\subfigure[]{\includegraphics[width=1\linewidth,left]{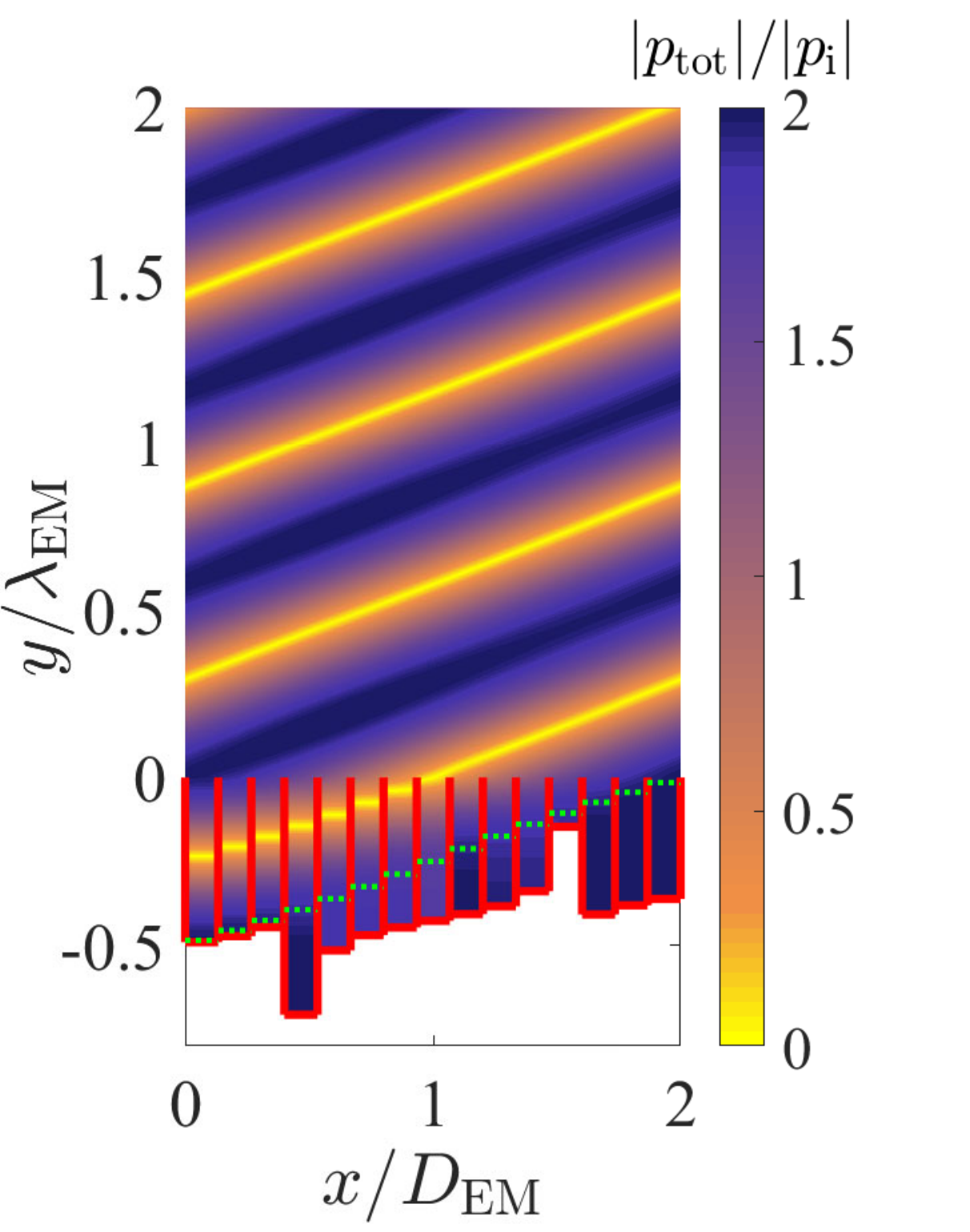}\label{fig:AnomalousReflectorA}}
\endminipage\hfill\minipage{0.47\columnwidth}
	\subfigure[]{\includegraphics[width=1\linewidth,left]{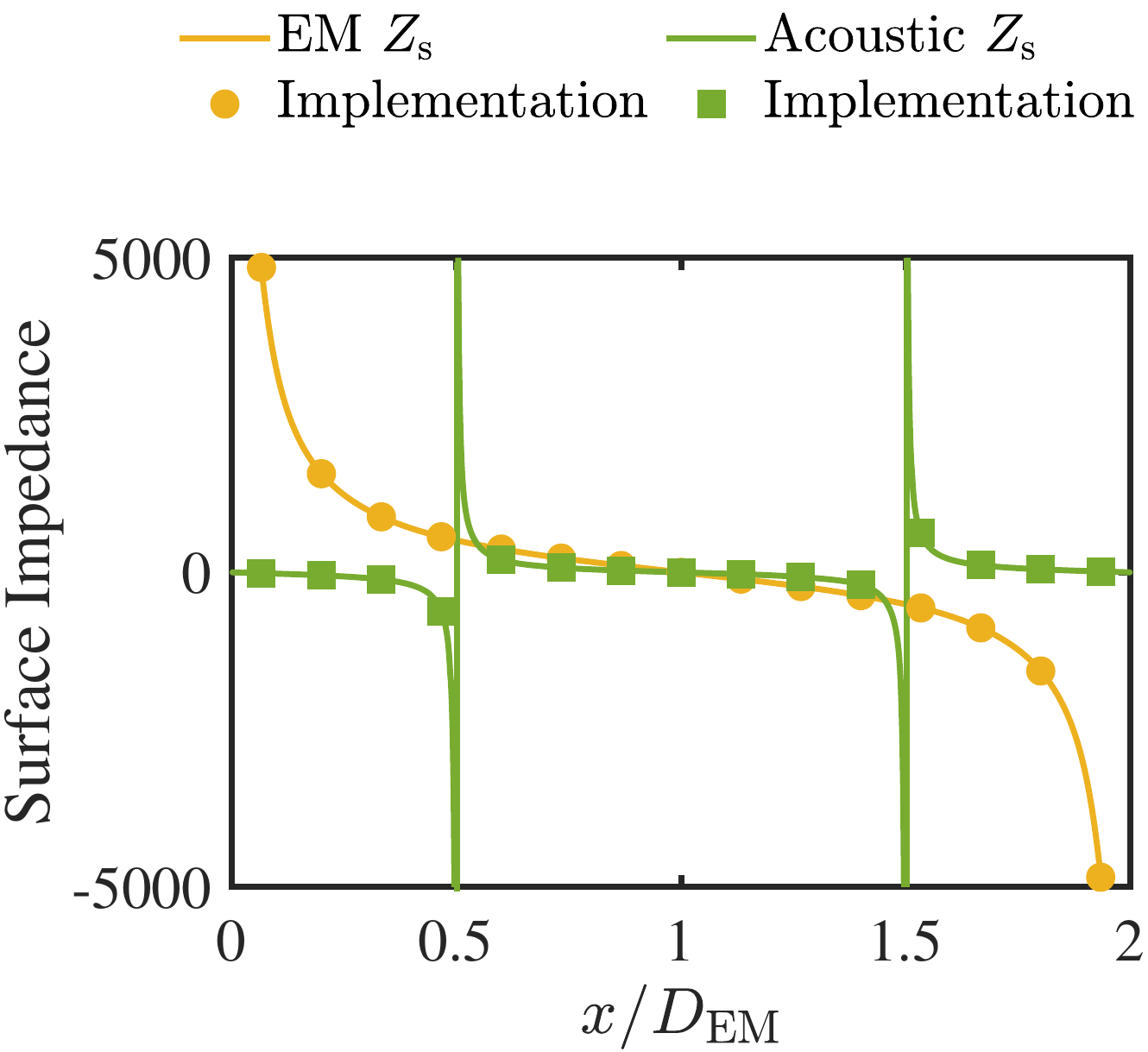}\label{fig:AnomalousReflectorC}}
	\subfigure[]{\includegraphics[width=1\linewidth,left]{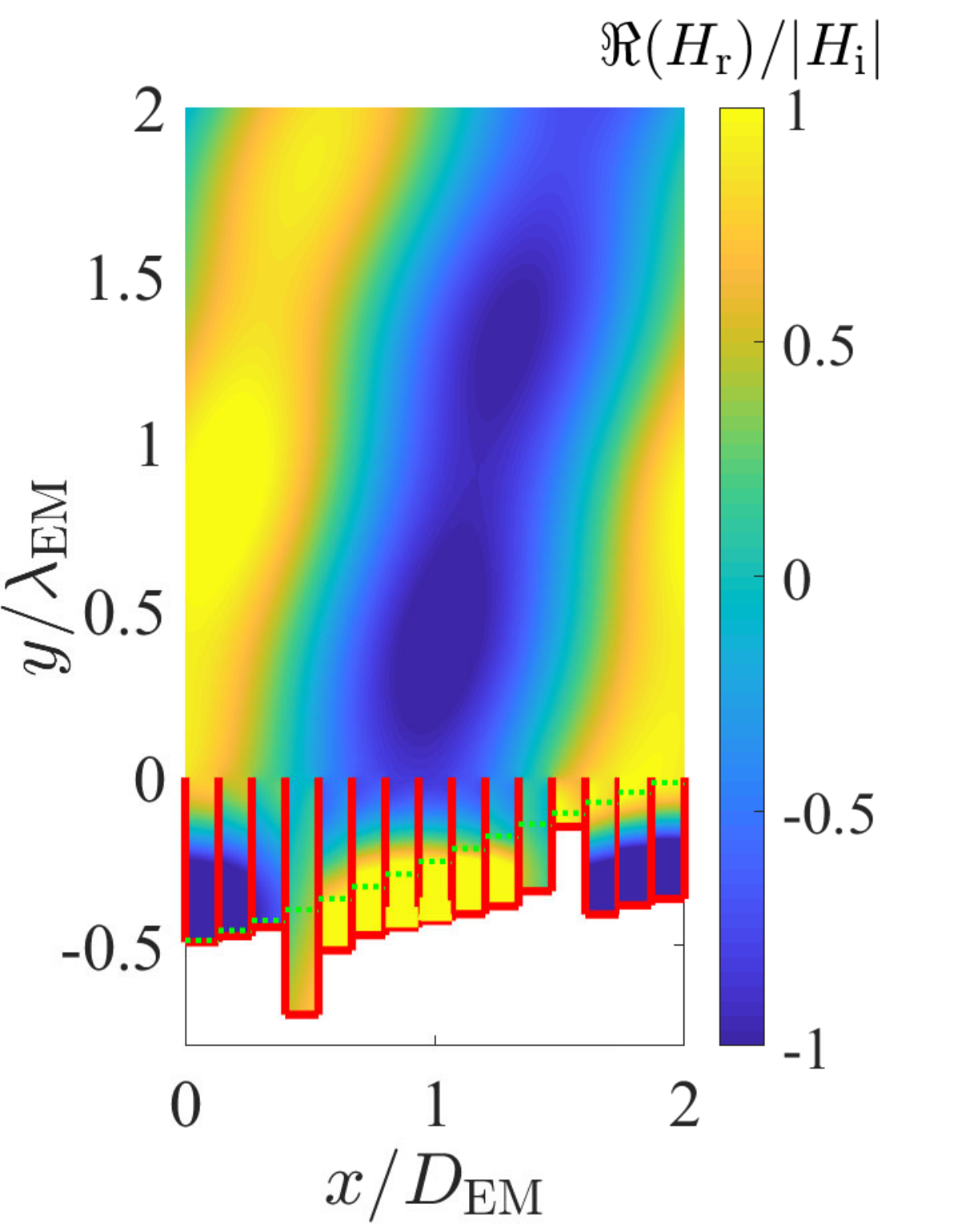}\label{fig:AnomalousReflectorA}}
\endminipage\hfill
	\caption{Multidisciplinary retroreflector for different angles. (a) Schematic representation of the metamirror. (b) Total pressure field calculated with a numerical simulation for acoustic waves. Red lines denote the hard boundary conditions used for simulating the groves. Green dotted lines represent the transition between empty grove and the solid material. (c) Surface impedance for electromagnetic (EM) and acoustic metamirrors.  (d) Reflected magnetic field calculated with a numerical simulation for EM waves. Red lines denote the PEC boundary conditions used for simulating the grove walls. Green dotted lines represent the transition between empty grove volume and the filling dielectric material.}
	\label{fig:Case3}
\end{figure}

\textit{\underline{Multidisciplinary retroreflector for different angles}:} Here, we propose an implementation of flat retroreflectors able to work at different angles for acoustics and electromagnetic waves. The incident angle of electromagnetic illumination, $\theta_{\rm EM}$, will determine the periodicity for electromagnetic waves $D_{\rm EM}=\lambda_{\rm EM}/2\sin\theta_{\rm EM}$. Then, we define the periodicity for acoustic waves to be $m$-times bigger than the periodicity for electromagnetic waves $D_{\rm ac}=\lambda_{\rm ac}/2\sin\theta_{\rm ac}=m D_{\rm EM}$. Free  choice of the coefficient $m$ allows us to control the operating incident angle for acoustic waves as  $\theta_{\rm ac}=\arcsin{(\lambda_{\rm ac}\sin\theta_{\rm EM}/m \lambda_{\rm EM})}$. In our example, we will assume that $\lambda_{\rm ac}=\lambda_{\rm EM}$ by choosing the operation frequencies equal to $f_{\rm EM}=3$~GHz and  $f_{\rm ac}=3430$~Hz. As another example, if we choose $m=2$ and $\theta_{\rm EM}=80^\circ$, then $\theta_{\rm ac}=\arcsin{(\sin\theta_{\rm EM}/2)}=29.5^\circ$. 

The next step in the design is to implement the required surface impedance for both acoustic and electromagnetic scenarios, see  Fig.~\ref{fig:Case3}(c). The main difference with the previous examples is that in this case there is no relation between the acoustic and electromagnetic surface impedances. For this reason, we need to use meta-atoms that offer  independent control of both impedances. In this example, we choose  partially filled close-ended groves as meta-atoms. We chose a material used for filling the tube with the relative permittivity $\varepsilon_2=2$. To systematically design the meta-atoms, we start implementing the acoustic response according to $Z_{\rm s}^{\rm ac}=-jZ_0^{\rm ac}\cot(k_{\rm ac}\ell_1)$. The depths of the empty volume of each grove, $\ell_1$,  equal 48.5~mm, 45.6~mm, 42.6~mm, 39.4~mm, 36.1~mm, 32.5~mm, 28.8~mm, 25~mm, 21.2~mm, 17.5~mm, 13.9~mm, 10.6~mm, 7.4~mm, 4.4~mm, and 1.5~mm. Using Eq.~\eqref{eq7} we can calculate the required length of the dielectric filling $\ell_2$. The found values of  $\ell_2$ are 0.6~mm, 1.7~mm, 2~mm, 31.3~mm, 15.4~mm, 14.5~mm, 15.9~mm, 17.7~mm, 19.5~mm, 20.9~mm, 20~mm, 4.1~mm, 33.4~mm, 33.7~mm, and 34.7~mm. In Fig.~\ref{fig:Case3}(b), a comparison between the required surface impedances and the impedances implemented with the actual meta-atoms is presented. Figures~\ref{fig:Case3}(c) and (d) show the results of a numerical simulation of the structure response for acoustic and electromagnetic waves where we can see standing waves generated by the incident and reflected waves, with different operational angles for acoustic and electromagnetic waves.

\begin{figure}[h]
\minipage{0.47\columnwidth}
	\subfigure[]{\includegraphics[width=1\linewidth,left]{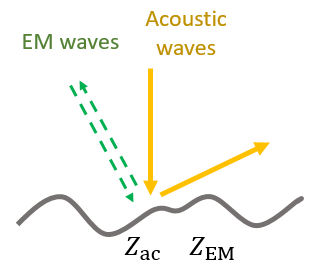}\label{fig:Fig9A}}
	\subfigure[]{\includegraphics[width=1\linewidth,left]{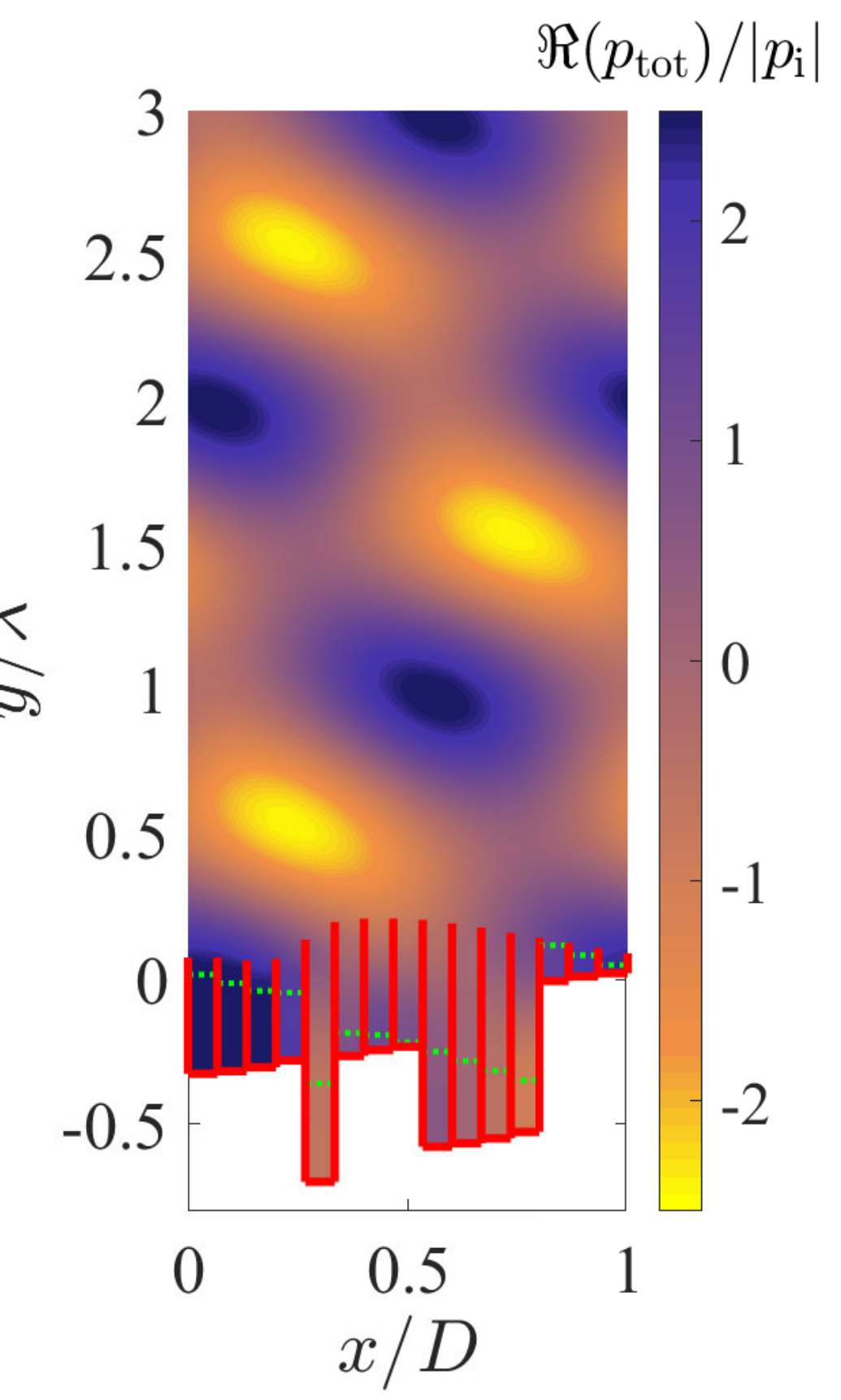}\label{fig:Fig9B}}
\endminipage\hfill\minipage{0.47\columnwidth}
	\subfigure[]{\includegraphics[width=1\linewidth,left]{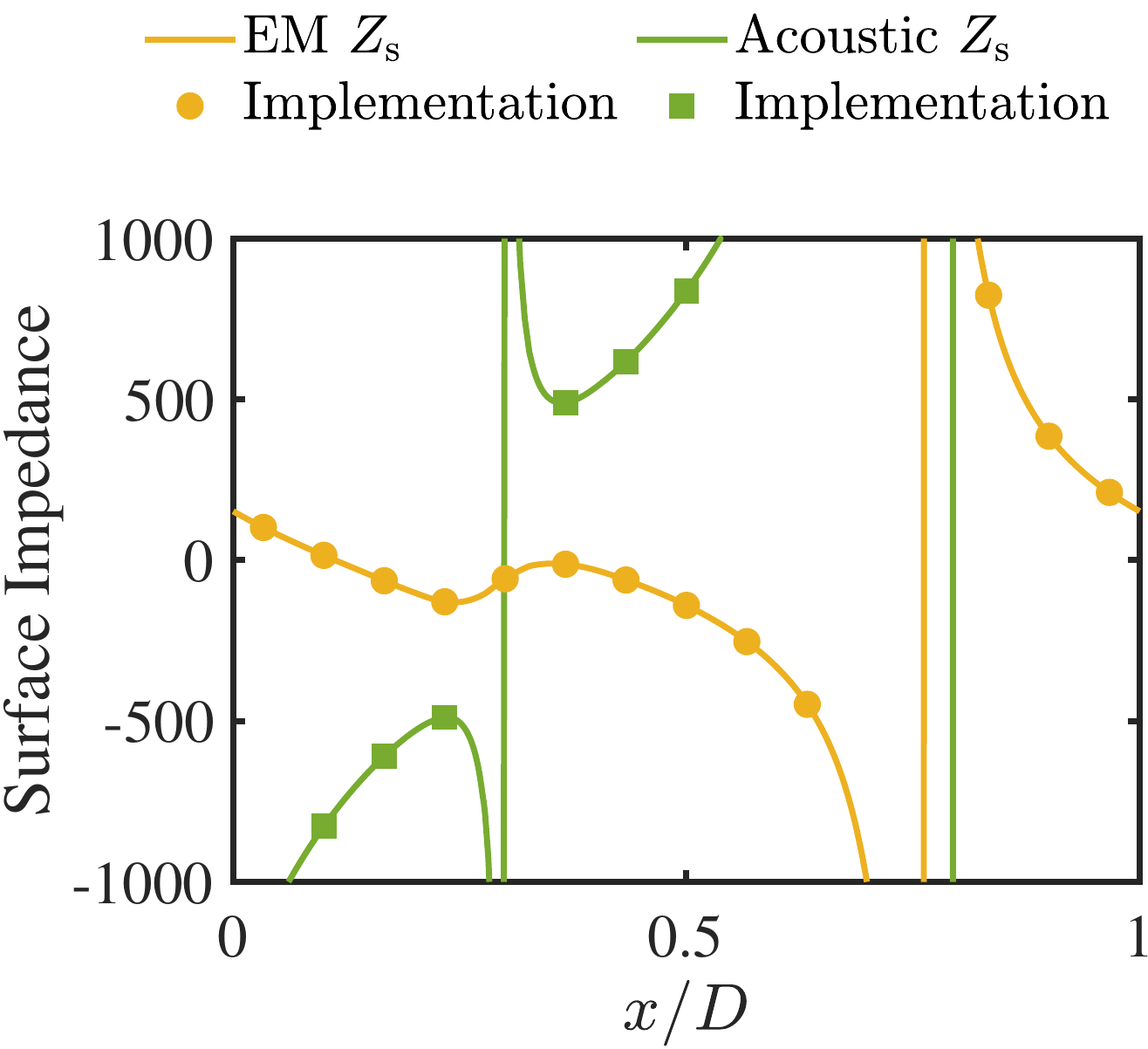}\label{fig:Fig9C}}
	\subfigure[]{\includegraphics[width=1\linewidth,left]{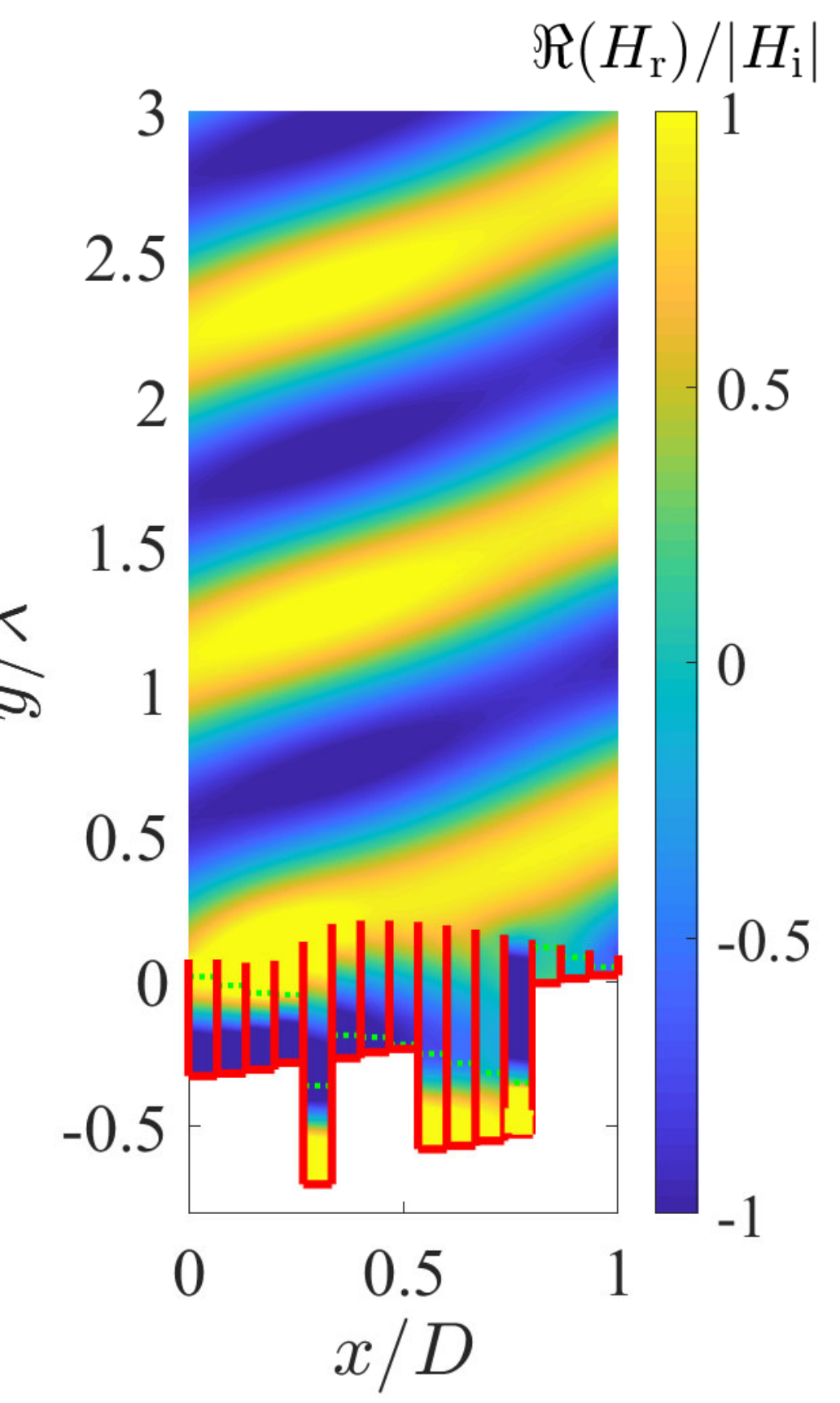}\label{fig:Fig9D}}
\endminipage\hfill
	\caption{Acoustic anomalous reflector and electromagnetic retroreflector. (a) Schematic representation of the metamirror. (b) Total pressure field calculated with a numerical simulation for acoustic waves. Red lines show  the hard boundary conditions used for simulating the groves. Green dotted lines indicate the transition between empty grove volume and the solid filling material. (c) Surface impedance for electromagnetic (EM) and acoustic metamirror.  (d) Reflected magnetic field calculated with a numerical simulation for EM waves. Red lines denote the PEC boundaries used for simulating the grove walls. Green dotted lines represent the transition between empty grove volume and the dielectric filling material. }
	\label{fig:Case3}
\end{figure}

\textit{\underline{Acoustic anomalous reflector and EM retroreflector}:} In this last example, we propose and study a device that acts a a retroreflector for electromagnetic waves and as an anomalous reflector for acoustic waves. Since  the electromagnetic retroreflector can be implemented with any surface profile, we start the design with the definition of the tangential to the power flow surface for the acoustic anomalous reflector. Using the theoretical approach described in Section~IIC, we find the surface profile, $y_{\rm c}=f(x)$, and calculate the surface acoustic impedance $Z_{\rm s}^{\rm ac}$. For the incident angle $\theta_i^{\rm ac}=0^\circ$ and the reflected angle $\theta_r^{\rm ac}=60^\circ$, the acoustic impedance is represented in Fig.~\ref{fig:Fig9C}. In this example, for the sake of simplicity, we chose the incident angle of the electromagnetic retroreflector, $\theta_i^{\rm EM}$, that corresponds to the same surface period as the acoustic anomalous reflector, $D_{\rm ac}=D_{\rm EM}$. Choosing the operation frequencies equal to $f_{\rm EM}=3$~GHz and  $f_{\rm ac}=3430$~Hz, the angle of incidence for the electromagnetic retroreflector is $\theta_i^{\rm EM}=25.65^\circ$. The electromagnetic surface impedance $Z_{\rm s}^{\rm EM}$ for the surface profile defined by $y_{\rm c}$ is represented in Fig.~\ref{fig:Fig9C}. As it was mentioned before, if the electromagnetic and acoustic periods are not equal, the period of the structure is chosen as their least common multiple.

Finally, once we know a suitable surface profile and the corresponding surface impedances, we design  meta-atoms that implement the surface impedance for both scenarios. Because this device requires independent control of electromagnetic and acoustic impedances, we  choose  as meta-atoms close-ended groves  partially filled with  a dielectric with  relative  permittivity $\varepsilon_2=2$. Using the analytical formulas presented at the beginning of this  section, the  lengths  of the  empty  portion of the groves, $\ell_1$, that produce the desired acoustic response will be 5.9~mm, 7.8~mm, 10~mm, 11.7~mm, 49.5~mm, 38.3~mm, 40.1~mm, 42.2~mm, 44.2~mm, 46~mm, 47.6~mm, 49.2~mm, 0.8~mm, 2.4~mm, and 4.1~mm. Then, using Eq.~\eqref{eq7}, we calculate the required lengths of the dielectric filling $\ell_2$. The found values of  $\ell_2$ are 34.2~mm, 30.3~mm, 26.4~mm, 23.4~mm, 34~mm, 7.9~mm, 5.2~mm, 1.5~mm, 32.8~mm, 28.4~mm, 23.4~mm, 17.8~mm, 12.3~mm, 7.3~mm, and 2.8~mm. Figures~\ref{fig:Fig9B} and \ref{fig:Fig9D} show the results of numerical simulations for acoustic and electromagnetic waves. In Fig.~\ref{fig:Fig9B}, the total acoustic field is represented and we can recognize the typical interference pattern generated by the incident wave and the reflected wave. Figure~\ref{fig:Fig9D} shows the reflected magnetic field where we can see that the energy is sent back into the illumination direction.


\section {Conclusions}

Multidisciplinary analysis of power flow-conformal metamirros has been performed from both acoustic and electromagnetic points of view. Based on the revealed analogy we have found a possibility to create metamirrors which operate as various anomalous reflectors both for electromagnetic and acoustic waves at the same tine. We propose three simple meta-atom topologies that allow us to engineer and control the electromagnetic and the acoustic responses at will. 
The considered example functionalities include high-efficient (theoretically perfect) retroreflection and anomalous reflection for both waves.  The theory and design approach  can be expanded to more complex transformations such as, for instance, beam splitters for waves of one nature and retroreflectors for  waves of the other nature. Thanks to the analytical formulation, the design process is straightforward and numerical optimizations are not required. 
This multidisciplinary approach opens a 
path for the design of multiphysics devices that integrate functionalities for both waves. Finally we note that the proposed topology of dual-physics meta-atoms allow electrical tunability, for example, using switches to control metal wire grids or using tunable materials as filling for the groves or tubes. 

\section*{Acknowledgment}
	
	This work received funding from the  the Academy of Finland under the grant project 309421.

\bibliography{references}

\begin{thebibliography}{16}%
\makeatletter
\providecommand \@ifxundefined [1]{%
 \@ifx{#1\undefined}
}%
\providecommand \@ifnum [1]{%
 \ifnum #1\expandafter \@firstoftwo
 \else \expandafter \@secondoftwo
 \fi
}%
\providecommand \@ifx [1]{%
 \ifx #1\expandafter \@firstoftwo
 \else \expandafter \@secondoftwo
 \fi
}%
\providecommand \natexlab [1]{#1}%
\providecommand \enquote  [1]{``#1''}%
\providecommand \bibnamefont  [1]{#1}%
\providecommand \bibfnamefont [1]{#1}%
\providecommand \citenamefont [1]{#1}%
\providecommand \href@noop [0]{\@secondoftwo}%
\providecommand \href [0]{\begingroup \@sanitize@url \@href}%
\providecommand \@href[1]{\@@startlink{#1}\@@href}%
\providecommand \@@href[1]{\endgroup#1\@@endlink}%
\providecommand \@sanitize@url [0]{\catcode `\\12\catcode `\$12\catcode
  `\&12\catcode `\#12\catcode `\^12\catcode `\_12\catcode `\%12\relax}%
\providecommand \@@startlink[1]{}%
\providecommand \@@endlink[0]{}%
\providecommand \url  [0]{\begingroup\@sanitize@url \@url }%
\providecommand \@url [1]{\endgroup\@href {#1}{\urlprefix }}%
\providecommand \urlprefix  [0]{URL }%
\providecommand \Eprint [0]{\href }%
\providecommand \doibase [0]{http://dx.doi.org/}%
\providecommand \selectlanguage [0]{\@gobble}%
\providecommand \bibinfo  [0]{\@secondoftwo}%
\providecommand \bibfield  [0]{\@secondoftwo}%
\providecommand \translation [1]{[#1]}%
\providecommand \BibitemOpen [0]{}%
\providecommand \bibitemStop [0]{}%
\providecommand \bibitemNoStop [0]{.\EOS\space}%
\providecommand \EOS [0]{\spacefactor3000\relax}%
\providecommand \BibitemShut  [1]{\csname bibitem#1\endcsname}%
\let\auto@bib@innerbib\@empty
\bibitem [{\citenamefont {Assouar}\ \emph {et~al.}(2018)\citenamefont
  {Assouar}, \citenamefont {Liang}, \citenamefont {Wu}, \citenamefont {Li},
  \citenamefont {Cheng},\ and\ \citenamefont {Jing}}]{Review_AC}%
  \BibitemOpen
  \bibfield  {author} {\bibinfo {author} {\bibfnamefont {B.}~\bibnamefont
  {Assouar}}, \bibinfo {author} {\bibfnamefont {B.}~\bibnamefont {Liang}},
  \bibinfo {author} {\bibfnamefont {Y.}~\bibnamefont {Wu}}, \bibinfo {author}
  {\bibfnamefont {Y.}~\bibnamefont {Li}}, \bibinfo {author} {\bibfnamefont
  {J.-C.}\ \bibnamefont {Cheng}}, \ and\ \bibinfo {author} {\bibfnamefont
  {Y.}~\bibnamefont {Jing}},\ }\href@noop {} {\bibfield  {journal} {\bibinfo
  {journal} {Nature Reviews Materials}\ }\textbf {\bibinfo {volume} {3}},\
  \bibinfo {pages} {460} (\bibinfo {year} {2018})}\BibitemShut {NoStop}%
\bibitem [{\citenamefont {Glybovski}\ \emph {et~al.}(2016)\citenamefont
  {Glybovski}, \citenamefont {Tretyakov}, \citenamefont {Belov}, \citenamefont
  {Kivshar},\ and\ \citenamefont {Simovski}}]{Review_EM}%
  \BibitemOpen
  \bibfield  {author} {\bibinfo {author} {\bibfnamefont {S.}~\bibnamefont
  {Glybovski}}, \bibinfo {author} {\bibfnamefont {S.}~\bibnamefont
  {Tretyakov}}, \bibinfo {author} {\bibfnamefont {P.}~\bibnamefont {Belov}},
  \bibinfo {author} {\bibfnamefont {Y.}~\bibnamefont {Kivshar}}, \ and\
  \bibinfo {author} {\bibfnamefont {C.}~\bibnamefont {Simovski}},\ }\href@noop
  {} {\bibfield  {journal} {\bibinfo  {journal} {Physics Reports}\ }\textbf
  {\bibinfo {volume} {634}},\ \bibinfo {pages} {1} (\bibinfo {year}
  {2016})}\BibitemShut {NoStop}%
\bibitem [{\citenamefont {Quevedo-Teruel}(2019)}]{Quevedo_Roadmap_2019}%
  \BibitemOpen
  \bibfield  {author} {\bibinfo {author} {\bibfnamefont {O.}~\bibnamefont
  {Quevedo-Teruel}},\ }\href@noop {} {\bibfield  {journal} {\bibinfo  {journal}
  {Journal of Optics}\ }\textbf {\bibinfo {volume} {21}} (\bibinfo {year}
  {2019})}\BibitemShut {NoStop}%
\bibitem [{\citenamefont {Estakhri}\ and\ \citenamefont
  {Alu}(2016)}]{Estakhri_Wavefront_2016}%
  \BibitemOpen
  \bibfield  {author} {\bibinfo {author} {\bibfnamefont {N.}~\bibnamefont
  {Estakhri}}\ and\ \bibinfo {author} {\bibfnamefont {A.}~\bibnamefont {Alu}},\
  }\href@noop {} {\bibfield  {journal} {\bibinfo  {journal} {Physical Review
  X}\ }\textbf {\bibinfo {volume} {6}},\ \bibinfo {pages} {041008} (\bibinfo
  {year} {2016})}\BibitemShut {NoStop}%
\bibitem [{\citenamefont {D\'{i}az-Rubio}\ \emph {et~al.}(2017)\citenamefont
  {D\'{i}az-Rubio}, \citenamefont {Asadchy}, \citenamefont {Elsakka},\ and\
  \citenamefont {Tretyakov}}]{Diaz_From_2017}%
  \BibitemOpen
  \bibfield  {author} {\bibinfo {author} {\bibfnamefont {A.}~\bibnamefont
  {D\'{i}az-Rubio}}, \bibinfo {author} {\bibfnamefont {V.}~\bibnamefont
  {Asadchy}}, \bibinfo {author} {\bibfnamefont {A.}~\bibnamefont {Elsakka}}, \
  and\ \bibinfo {author} {\bibfnamefont {S.}~\bibnamefont {Tretyakov}},\
  }\href@noop {} {\bibfield  {journal} {\bibinfo  {journal} {Science Advances}\
  }\textbf {\bibinfo {volume} {3}},\ \bibinfo {pages} {e1602714} (\bibinfo
  {year} {2017})}\BibitemShut {NoStop}%
\bibitem [{\citenamefont {Epstein}\ and\ \citenamefont
  {Rabinovich}(2017)}]{Epstein_Unveiling_2017}%
  \BibitemOpen
  \bibfield  {author} {\bibinfo {author} {\bibfnamefont {A.}~\bibnamefont
  {Epstein}}\ and\ \bibinfo {author} {\bibfnamefont {O.}~\bibnamefont
  {Rabinovich}},\ }\href@noop {} {\bibfield  {journal} {\bibinfo  {journal}
  {Physical Review Applied}\ }\textbf {\bibinfo {volume} {8}},\ \bibinfo
  {pages} {054037} (\bibinfo {year} {2017})}\BibitemShut {NoStop}%
\bibitem [{\citenamefont {Ra’di}\ \emph {et~al.}(2018)\citenamefont
  {Ra’di}, \citenamefont {Sounas},\ and\ \citenamefont
  {Alu}}]{Radi_Metagrating_2018}%
  \BibitemOpen
  \bibfield  {author} {\bibinfo {author} {\bibfnamefont {Y.}~\bibnamefont
  {Ra’di}}, \bibinfo {author} {\bibfnamefont {D.~L.}\ \bibnamefont {Sounas}},
  \ and\ \bibinfo {author} {\bibfnamefont {A.}~\bibnamefont {Alu}},\
  }\href@noop {} {\bibfield  {journal} {\bibinfo  {journal} {Physical Review
  Letters}\ }\textbf {\bibinfo {volume} {119}} (\bibinfo {year}
  {2018})}\BibitemShut {NoStop}%
\bibitem [{\citenamefont {Kwon}(2018)}]{DoHoon_Lossless_2018}%
  \BibitemOpen
  \bibfield  {author} {\bibinfo {author} {\bibfnamefont {D.}~\bibnamefont
  {Kwon}},\ }\href@noop {} {\bibfield  {journal} {\bibinfo  {journal} {IEEE
  Antennas and Wireless Propagation Letters}\ }\textbf {\bibinfo {volume} {17}}
  (\bibinfo {year} {2018})}\BibitemShut {NoStop}%
\bibitem [{\citenamefont {Asadchy}\ \emph {et~al.}(2017)\citenamefont
  {Asadchy}, \citenamefont {D\'{i}az-Rubio}, \citenamefont {Tcvetkova},
  \citenamefont {Kwon}, \citenamefont {Elsakka}, \citenamefont {Albooyeh},\
  and\ \citenamefont {Tretyakov}}]{Asadchy_Flat_2017}%
  \BibitemOpen
  \bibfield  {author} {\bibinfo {author} {\bibfnamefont {V.}~\bibnamefont
  {Asadchy}}, \bibinfo {author} {\bibfnamefont {A.}~\bibnamefont
  {D\'{i}az-Rubio}}, \bibinfo {author} {\bibfnamefont {S.}~\bibnamefont
  {Tcvetkova}}, \bibinfo {author} {\bibfnamefont {D.}~\bibnamefont {Kwon}},
  \bibinfo {author} {\bibfnamefont {A.}~\bibnamefont {Elsakka}}, \bibinfo
  {author} {\bibfnamefont {M.}~\bibnamefont {Albooyeh}}, \ and\ \bibinfo
  {author} {\bibfnamefont {S.}~\bibnamefont {Tretyakov}},\ }\href@noop {}
  {\bibfield  {journal} {\bibinfo  {journal} {7}\ }\textbf {\bibinfo {volume}
  {3}} (\bibinfo {year} {2017})}\BibitemShut {NoStop}%
\bibitem [{\citenamefont {D\'{i}az-Rubio}\ and\ \citenamefont
  {Tretyakov}(2017)}]{Diaz_Acoustic_2017}%
  \BibitemOpen
  \bibfield  {author} {\bibinfo {author} {\bibfnamefont {A.}~\bibnamefont
  {D\'{i}az-Rubio}}\ and\ \bibinfo {author} {\bibfnamefont {S.}~\bibnamefont
  {Tretyakov}},\ }\href@noop {} {\bibfield  {journal} {\bibinfo  {journal}
  {Physical Review B}\ }\textbf {\bibinfo {volume} {96}},\ \bibinfo {pages}
  {125409} (\bibinfo {year} {2017})}\BibitemShut {NoStop}%
\bibitem [{\citenamefont {Torrent}(2018)}]{Torrent_2018}%
  \BibitemOpen
  \bibfield  {author} {\bibinfo {author} {\bibfnamefont {D.}~\bibnamefont
  {Torrent}},\ }\href@noop {} {\bibfield  {journal} {\bibinfo  {journal}
  {Physical Review B}\ }\textbf {\bibinfo {volume} {98}},\ \bibinfo {pages}
  {060101(R)} (\bibinfo {year} {2018})}\BibitemShut {NoStop}%
\bibitem [{\citenamefont {D\'{i}az-Rubio}\ \emph {et~al.}(2019)\citenamefont
  {D\'{i}az-Rubio}, \citenamefont {Li}, \citenamefont {Shen}, \citenamefont
  {Cummer},\ and\ \citenamefont {Tretyakov}}]{Diaz_Power_2019}%
  \BibitemOpen
  \bibfield  {author} {\bibinfo {author} {\bibfnamefont {A.}~\bibnamefont
  {D\'{i}az-Rubio}}, \bibinfo {author} {\bibfnamefont {J.}~\bibnamefont {Li}},
  \bibinfo {author} {\bibfnamefont {C.}~\bibnamefont {Shen}}, \bibinfo {author}
  {\bibfnamefont {S.}~\bibnamefont {Cummer}}, \ and\ \bibinfo {author}
  {\bibfnamefont {S.}~\bibnamefont {Tretyakov}},\ }\href@noop {} {\bibfield
  {journal} {\bibinfo  {journal} {Science Advances}\ }\textbf {\bibinfo
  {volume} {5}},\ \bibinfo {pages} {eaau7288} (\bibinfo {year}
  {2019})}\BibitemShut {NoStop}%
\bibitem [{\citenamefont {Shen}\ \emph {et~al.}(2018)\citenamefont {Shen},
  \citenamefont {D\'{i}az-Rubio}, \citenamefont {Li},\ and\ \citenamefont
  {Cummer}}]{Shen_Surface_2018}%
  \BibitemOpen
  \bibfield  {author} {\bibinfo {author} {\bibfnamefont {C.}~\bibnamefont
  {Shen}}, \bibinfo {author} {\bibfnamefont {A.}~\bibnamefont
  {D\'{i}az-Rubio}}, \bibinfo {author} {\bibfnamefont {J.}~\bibnamefont {Li}},
  \ and\ \bibinfo {author} {\bibfnamefont {S.}~\bibnamefont {Cummer}},\
  }\href@noop {} {\bibfield  {journal} {\bibinfo  {journal} {Applied Physics
  Letters}\ }\textbf {\bibinfo {volume} {112}},\ \bibinfo {pages} {183503}
  (\bibinfo {year} {2018})}\BibitemShut {NoStop}%
\bibitem [{\citenamefont {Popov}\ \emph {et~al.}(2018)\citenamefont {Popov},
  \citenamefont {Boust},\ and\ \citenamefont
  {Burokur}}]{Popov_Controlling_2018}%
  \BibitemOpen
  \bibfield  {author} {\bibinfo {author} {\bibfnamefont {V.}~\bibnamefont
  {Popov}}, \bibinfo {author} {\bibfnamefont {F.}~\bibnamefont {Boust}}, \ and\
  \bibinfo {author} {\bibfnamefont {S.~N.}\ \bibnamefont {Burokur}},\
  }\href@noop {} {\bibfield  {journal} {\bibinfo  {journal} {Physical Review
  Applied}\ }\textbf {\bibinfo {volume} {10}},\ \bibinfo {pages} {011002}
  (\bibinfo {year} {2018})}\BibitemShut {NoStop}%
\bibitem [{\citenamefont {Carbonell}\ \emph {et~al.}(2011)\citenamefont
  {Carbonell}, \citenamefont {Torrent}, \citenamefont {D\'{i}az-Rubio},\ and\
  \citenamefont {S\'{a}nchez-Dehesa}}]{Carbonell_Multidisciplinary_2011}%
  \BibitemOpen
  \bibfield  {author} {\bibinfo {author} {\bibfnamefont {J.}~\bibnamefont
  {Carbonell}}, \bibinfo {author} {\bibfnamefont {D.}~\bibnamefont {Torrent}},
  \bibinfo {author} {\bibfnamefont {A.}~\bibnamefont {D\'{i}az-Rubio}}, \ and\
  \bibinfo {author} {\bibfnamefont {J.}~\bibnamefont {S\'{a}nchez-Dehesa}},\
  }\href@noop {} {\bibfield  {journal} {\bibinfo  {journal} {New Journal of
  Physics}\ }\textbf {\bibinfo {volume} {13}},\ \bibinfo {pages} {103034}
  (\bibinfo {year} {2011})}\BibitemShut {NoStop}%
\bibitem [{\citenamefont {Tretyakov}(2003)}]{modeboo}%
  \BibitemOpen
  \bibfield  {author} {\bibinfo {author} {\bibfnamefont {S.}~\bibnamefont
  {Tretyakov}},\ }\href@noop {} {\emph {\bibinfo {title} {Analytical Modeling
  in Applied Electromagnetics}}}\ (\bibinfo  {publisher} {Norwood, MA: Artech
  House},\ \bibinfo {year} {2003})\BibitemShut {NoStop}%
\end{thebibliography}%

\end{document}